\documentclass[11pt, oneside]{article}   	
\pdfoutput=1
\usepackage{jheppub}
\usepackage{amsmath, amsthm, graphicx,amssymb,subfigure}
\usepackage[all]{xy}
\usepackage{float}
\usepackage{color}
\usepackage[position=below]{caption}

\usepackage{cancel}
\usepackage{ulem}
\usepackage[numbers]{natbib}
\newcommand{\nn}{\nonumber}
\newcommand{\bxo}{\bar{X}^{\rho \dot{\rho}}}
\newcommand{\xo}{X_{\rho \dot{\rho}}}
\newcommand{\xon}{X_{n\rho \dot{\rho}}}

\newcommand{\tr}{\operatorname{tr}}
\newcommand{\Tr}{\operatorname{Tr}}
\newcommand{\adj}{\operatorname{ad}}

\newcommand{\iDelta}{\mathit{\Delta}}
\allowdisplaybreaks

\title{The Flavoured BFSS Model at High Temperature}

\author[a]{Yuhma Asano,}
\author[a]{Veselin G. Filev,}
\author[a,b]{Samuel Kov\'a\v{c}ik,}
\author[a]{Denjoe O'Connor,}
\affiliation[a]{School of Theoretical Physics,\\ 
       Dublin Institute for Advanced Studies, \\
       10 Burlington Road, 
       Dublin 4, Ireland.}
\affiliation[b]{Faculty of Mathematics, Physics and Informatics, \\
  Comenius University Bratislava, \\
  Mlynsk\'a dolina, Bratislava,\\
  842 48, Slovakia.}
\emailAdd{yuhma@stp.dias.ie}
\emailAdd{vfilev@stp.dias.ie}
\emailAdd{samuel.kovacik@fmph.uniba.sk}
\emailAdd{denjoe@stp.dias.ie}

\abstract{We study the high-temperature series expansion of the
  Berkooz-Douglas matrix model, which describes the D0/D4--brane
  system. At high temperature the model is weakly coupled and we
  develop the series to second order. We check our results against the
  high-temperature regime of the bosonic model (without fermions) and
  find excellent agreement. We track the temperature dependence of the
  bosonic model and find backreaction of the fundamental
  fields lifts the zero-temperature adjoint mass degeneracy.
  In the low-temperature phase the system is well
  described by a gaussian model with three masses $m^t_A=1.964 \pm
  0.003$, $m^l_A=2.001 \pm 0.003$ and $m_f=1.463 \pm 0.001$, the
  adjoint longitudinal and transverse masses and the mass of the
  fundamental fields respectively. 
}

\begin{document}

\maketitle

\section{Introduction}
The Berkooz-Douglas model (BD model) \cite{Berkooz:1996is} was
introduced as a non-perturbative formulation of M-theory in the
presence of a background of longitudinal M5-branes with the M2-brane
quantised in light-cone gauge.  Its action is written as that of the
BFSS model \cite{Banks:1996vh} with additional fundamental
hypermultiplets to describe the M5-branes. The BFSS model can also be
viewed as a many-body system of D0-branes of the IIA superstring. In
this framework the BD model is a D0/D4 system with the massless case
being the D0/D4 intersection.  When the number of D0-branes far
exceeds that of the D4-branes the dynamics of the D0-branes is only
weakly affected by that of the D4-branes and is captured by the IIA
supergravity background holographically dual to the BFSS model. In
this context the D4-branes, representing the fundamental fields of the
BD model, are treated as Born-Infeld probe 4-branes. This holographic
set up is a tractable realisation of gauge/gravity duality with
flavour.

Both the BFSS model and the BD model are supersymmetric quantum
mechanical models with an $SU(N)$ gauge symmetry.  When they are put
in a thermal bath they become strongly coupled at low temperature. At
finite temperature their gravity duals involve a black hole whose
Hawking-temperature is that of the thermal bath. These duals can be
used to provide non-perturbative predictions at low temperature. The
BFSS and BD models can also be studied by the standard
non-perturbative field theory method of Monte Carlo simulation.  These
models therefore provide excellent candidates for testing
gauge/gravity duality non-perturbatively and in a broken supersymmetric setting.

There are now several non-perturbative studies of the BFSS
model~\cite{Anagnostopoulos:2007fw,Catterall:2008yz,Kadoh:2015mka,Filev:2015hia,Hanada:2016zxj} and several recent reviews\cite{Joseph:2015xwa,Hanada:2016jok,O'Connor:2016goh}. Also, the BD model was recently studied
non-perturbatively in \cite{Filev:2015cmz}.  In all
cases the predictions from the gauge/gravity duals were found to be in
excellent agreement with that of Monte Carlo simulations of 
the finite-temperature models.

The situation is conceptually simpler at high temperature as the dimensionless
inverse temperature, scaled in terms of the BD-coupling, provides a natural small
parameter for the model.  In this paper, we obtain the first two terms in the
high-temperature expansion of the BD model.

In the high-temperature limit only the bosonic Matsubara zero modes
survive and the resulting model is a pure potential.
This potential, which provides the non-perturbative aspect of
our high-temperature study, also plays a role in the ADHM
construction \cite{Tong:2005un}. We study the model for adjoint
matrix size $N$ between $4$ and $32$ for $N_f=1$ (with $N_f$ the number of
D4-branes) and for $N_f$ between $2$ and $16$ for $N$ from
$9$ to $20$. For $N_f\ge 2N$ we find that the system has difficulties
with ergodicity. In particular, for $N_f=2N$ and $N_f=2N+1$ the system
failed to thermalise satisfactorily. In contrast the system has no
difficulties for $N_f=2N-1$.
This condition is closely related to the singularity structure of
instanton moduli space, where irreducible $SU(N_f)$ instantons of Chern
number~$N$ exist only for $N\ge \frac{N_f}{2}$ \cite{Atiyah:1978wi,Nash:1991pb}.
The moduli space of such instantons is equivalent to the zero
locus of the potential with
$X^a=0$ and ${\cal D}^A=0$ (see equation (\ref{ADHM_D})).
This moduli space is in general singular
and non-singular only when this bound is satisfied.

There is also a natural $1+1$ dimensional analogue of the BD model,
which has ${\cal N}=4$ supersymmetry, associated with the D1/D5
system of \cite{VanRaamsdonk:2001cg}, whose BFSS relative was discussed in
\cite{Aharony:2004ig,Kawahara:2007fn,Mandal:2009vz}.
When the Euclidean finite-temperature version of
this $1+1$ dimensional quantum field theory
is considered on a torus with the spatial circle of period $\beta$ and euclidean
time\footnote{In this paragraph we avoid using $\beta$ for $1/T$ for
  simplicity of the comparison.}  of period $1/T$, then at high
temperature the fermions decouple and one is left with the purely
bosonic version of the BD model. We refer to this model as the bosonic
BD model and study the small period behaviour (equivalent for us to
our high-temperature regime) of the massless version of
this model as a check on our high-temperature series.
We find the high-temperature series results
are in excellent agreement with Monte Carlo simulations of the bosonic
BD model. By fitting the dependence, of the expectation values of our
observables, on the number of flavour multiplets, $N_f$,
we find that extrapolation, to $N_f=0$, agrees well with the
corresponding observables of the BFSS model.

As $\beta$, the inverse temperature, grows the bosonic BD model
undergoes a set of phase transitions. These are the phase transitions
of the bosonic BFSS model. We find the high-temperature series
expansion is valid up to $\beta\sim 1/2$, which is just below the
phase transition region. Above the transition the bosonic BD model is
well described by free massive fields, where the backreaction of the
fundamental fields has lifted the degeneracy of the longitudinal and
transverse masses.

The principal results of this paper are:
\begin{itemize}
\item{}We obtain expansions for observables of the BD model to second order in
  a high-temperature series.
\item{}We tabulate the coefficients of this expansion as functions of $N$ and $N_f$ in the range $4\le N\le 32$ and $1\le N_f\le 16$.
\item{}We measure the expectation values of the composite operator
  $\langle r^2\rangle_\text{bos}$, (see equation (\ref{defn_r2})), and the
  mass susceptibility $\langle {\mathcal C}^m\rangle_\text{bos}$,
  (see equation (\ref{condensate_susceptibility_bosonic})), of the bosonic BD model as a
  function of temperature down to zero and use it to check our
  coefficients for the high-temperature series of the full BD model.
\item{}We find that the fundamental fields of the bosonic BD model
  have mass $m_f=1.463 \pm 0.001$.
\item  We measure the backreacted mass of the longitudinal adjoint 
  scalars to be $m^l_A = 2.001 \pm 0.003$ and find that the transverse mass is largely
  unaffected by backreaction being $m^t_A=1.964 \pm 0.003$, which
  should be compared with the bosonic BFSS model, where the
  fields have mass $m_A= 1.965\pm0.007$.
\item  We use the measured masses to predict the zero-temperature values
  of our fundamental field observables $\langle r^2\rangle_\text{bos}$
  and mass susceptibility
  $\langle {\mathcal C}^m\rangle_\text{bos}$  and 
  find excellent agreement with direct measurements.
\end{itemize}

The paper is organised as follows: In section \ref{BD_model} we present the
finite-temperature BD model and describe our notation and observables.
In section \ref{High_Temp_Expans} we set up and implement the high-temperature
series expansion working to second order in the inverse temperature $\beta$.
Section~\ref{Numerical_sims} describes the dependence of our observables 
on the coefficients in the expansion, which must be determined by
numerical simulation of the zero-mode model. In Section~\ref{BosonicBDmodel}
we perform lattice simulations of the bosonic BD model and find excellent
agreement with the high-temperature expansion. We also find the low-temperature
phase of the model is well described by a system of gaussian quantum fields.
Section~\ref{Conclusions} gives our concluding remarks. There are two appendices;
appendix \ref{AppendixB} gives tables, for different $N$ and $N_f$,
of the coefficients determined non-perturbatively while 
appendix \ref{Predictions_for_observables_graphs} presents graphs
of predictions for the high-temperature behaviour of our
observables for the supersymmetric model.

\section{The Berkooz-Douglas Model}
\label{BD_model}
We begin by describing the field content of the model following
the notation used in \citep{VanRaamsdonk:2001cg}.  The action of the
BFSS model is given by
\begin{equation}\label{BFSS 10Mink}
S_{\rm BFSS} =\frac{1}{g^2}\int dt \sum_{i=1}^9\,{\rm Tr}\left\{\frac{1}{2}({\cal D}_0X^i)^2 +\frac{1}{4}[X^i,X^j]^2-\frac{i}{2}\Psi^T C_{10}\,\Gamma^0{\cal D}_0\Psi +\frac{1}{2}\Psi^T C_{10}\,\Gamma^i[X^i,\Psi]\right\}\ ,
\end{equation}
where ${\cal D}_0\;\cdot=\partial_t\;\cdot-i[A,\;\cdot\;]$, $\Psi$ is
a thirty-two component Majorana--Weyl spinor, $\Gamma^\mu$ are ten
dimensional gamma matrices and $C_{10}$ is the charge conjugation
matrix satisfying $C_{10} \Gamma^{\mu}C_{10}^{-1} =-{\Gamma^{\mu}}^T$.
The fields $X^i$ and $\Psi$ are in the adjoint representation of the gauge symmetry group $SU(N)$ and $A$ is the gauge field.

To describe the addition of the fundamental fields we break the $SO(9)$ vector
$X^i$ into an $SO(5)$ vector $X^a$ and an $SO(4)$ vector which 
we re-express as $X_{\rho\dot\rho}$  via\footnote{Here $X^8$ of \cite{VanRaamsdonk:2001cg} is replaced by $-X^8$.}   
\begin{align}
 X_{\rho\dot\rho}=
 \frac{i}{\sqrt{2}}\sum_{m=1}^{4}
 \sigma^{m}_{\;\; \rho\dot\rho}X^{10-m},
\end{align}
where $\sigma^4=-i\mathbf{1}_2$
and  $\sigma^A$'s ($A=1,2,3$) are the Pauli matrices.
The $X_{\rho\dot\rho}$ ($\rho,\dot \rho=1,2$)
are complex scalars which together transform as a real vector of
$SO(4)$ which satisfies the reality condition
$X_{\rho\dot\rho}=\varepsilon_{\rho\sigma}\varepsilon_{\dot\rho\dot\sigma}\bar X^{\sigma\dot\sigma}$. The indices $\rho$ and $\dot\rho$ are those of $SU(2)_R$ and $SU(2)_L$, respectively, where $SO(4)=SU(2)_L\times SU(2)_R$.

The nine BFSS scalar fields, $X^i$,
become $X^a$ ($a=1,\cdots ,5$) and $X_{\rho\dot\rho}$.
The sixteen adjoint fermions of the BFSS model become $\lambda_\rho$
and $\theta_{\dot\rho}$ with 
$\lambda_\rho$ being $SO(5,1)$ symplectic Majorana-Weyl spinors of
positive chirality and satisfying $\lambda_{\rho}=\varepsilon_{\rho\sigma} (\lambda^{c})^{\sigma}$ 
while $\theta_{\dot\rho}$ are symplectic Majorana-Weyl spinors of negative
chirality satisfying
$\theta_{\dot\rho}=-\varepsilon_{\dot\rho\dot\sigma} (\theta^{c})^{\dot\sigma}$. They 
combine together to form an $SO(9,1)$ Majorana-Weyl spinor in the
adjoint of $SU(N)$. 
This $SO(9)$ symmetry is recovered only if the fundamental
fields are turned off.

To describe the longitudinal M5-branes (or D4-branes), we
have $\Phi_{\rho}$ and $\chi$,  which transform in the fundamental
representations of both $SU(N)$ and the global
$SU(N_f)$ flavour symmetry. $\Phi_\rho$  are complex scalar fields 
with hermitian conjugates $\bar\Phi^\rho$,
and  $\chi$ is an $SO(5,1)$ spinor of negative chirality.

After rotating to imaginary time the Euclidean action describing the model
at finite temperature $T=\beta^{-1}$ becomes:
\begin{align}
 S=N\int_0^\beta d \tau  \, 
 &\Bigg[
 \Tr\left( \frac{1}{2}D_\tau X^a D_\tau X^a 
 + \frac{1}{2}D_\tau \bar{X}^{\rho \dot{\rho}} D_\tau X_{\rho \dot{\rho}} 
 + \frac{1}{2} \lambda^{\dagger\rho} D_\tau \lambda_\rho 
 + \frac{1}{2} \theta^{\dagger\dot{\rho}} D_\tau \theta_{\dot{\rho}} \right) 
 \nonumber\\
 &+\tr \left( D_\tau \bar{\Phi}^\rho D_\tau \Phi_\rho 
 + \chi^\dagger D_\tau \chi \right) 
 \nonumber\\
 &-\Tr \left( 
 \frac{1}{4} [X^a,X^b]^2
 + \frac{1}{2}[X^a, \bxo][X^a,\xo]
 \right)
 \nonumber\\
 &+\frac{1}{2}\Tr \sum_{A=1}^3\mathcal{D}^A\mathcal{D}^A
 +\tr \left( \bar{\Phi}^\rho (X^a-m^a)^2 \Phi _\rho \right) 
 \nonumber\\
 &-\Tr \left(
 -\frac{1}{2} \lambda^{\dagger\rho} \gamma^a [X^a , \lambda_\rho] 
 + \frac{1}{2}\theta^{\dagger\dot{\rho}} \gamma^a [X^a , \theta_{\dot{\rho}}] 
 - \sqrt{2}i \varepsilon^{\rho \sigma} \theta^{\dagger\dot{\rho}}
 [ X_{\sigma \dot{\rho}}, \lambda_\rho] 
 \right) \nonumber\\
 &-\tr \left( \chi^\dagger \gamma^a (X^a-m^a) \chi 
 + \sqrt{2}i \varepsilon^{\rho \sigma} \chi^\dagger \lambda_\rho \Phi_\sigma 
 + \sqrt{2}i \varepsilon_{\rho \sigma} 
 \bar{\Phi}^\rho \lambda^{\dagger\sigma} \chi \right) \,
 \Bigg] ,
 \label{action}
\end{align}
where 
\begin{align}
 &\mathcal{D}^A
 =\sigma^{A\; \sigma}_{\rho}\left(
 \frac{1}{2}[\bar{X}^{\rho \dot{\rho}}, X_{\sigma \dot{\rho}}]
 -\Phi_\sigma \bar\Phi^\rho
 \right)\, ,
\label{ADHM_D}
\end{align}
with $D_\tau$ the covariant derivative which, for the fields of the fundamental multiplet, $\Phi_\rho$ and $\chi$, acts as
$D_\tau\;\cdot=(\partial_\tau -iA)\;\cdot$ .
The trace of $SU(N)$ is written as $\Tr$ while that of $SU(N_f)$
is denoted by $\tr$.  The diagonal matrices, $m^a$,
correspond to the transverse positions of the D4-branes.

We fix the static gauge: $\partial_\tau A=0$, so the path integral
requires the
corresponding ghost fields $c$ and $\bar c$ with the ghost
term $N\int_0^\beta d\tau\, \Tr \partial_\tau \bar c D_\tau c$ 
added to the action \eqref{action}.

We will restrict our attention to $m^a=0$ so that the
D4-branes are attached to the D0-branes, 
and the strings between D0 and D4 are massless, i.e.~the
fundamental fields are massless. The factor of
$N$ in front of the integral in \eqref{action} is the remnant of 
the 't Hooft coupling $\lambda=g^2N$ which is kept fixed and
absorbed into $\tau$ and the fields with $\beta=\lambda^{1/3}/T$. Note that without loss of generality we can set
$\lambda=1$.

As discussed in the introduction, the BFSS model is the
matrix regularization of a supermembrane theory \cite{deWit:1988ig},
so the BFSS part of this model
can be also interpreted as M2-brane dynamics.
In this context the D4-branes lift to M5-branes and the model
can describe M2-branes ending on longitudinal M5-branes.

The BD model is a version of supersymmetric quantum mechanics
and could in principle be treated by Hamiltonian methods. The partition
function is then
\begin{equation}
Z=Tr({\rm e}^{-\beta H}) =\int[dX][d\lambda][d\theta][d\Phi][d\bar\Phi][d\chi][d\chi^\dag][dA]{\rm e}^{-S}
\label{PartitionFunction}
\end{equation}
with $Tr$ the trace over the Hilbert space of the Hamiltonian restricted to
its gauge invariant subspace and the action $S$, in the path integral, is given by equation (\ref{action}).

The measure in the path integral for the partition function
(\ref{PartitionFunction}) has a hidden dependence on temperature due
to the presence of the Van Vleck-Morette determinant
\cite{DeWitt:1992cy} in the definition of the path integral
measure. This determinant arises from the kinetic contribution to the
action (\ref{action}) which, as written, is temperature dependent. 
To remove the temperature dependence from the measure, we rescale the
variables in the original action (\ref{action}) so that the kinetic
terms, including the gauge potential, are independent of $\beta$.  For
this $\tau\rightarrow\beta\tau$,
$X^i\rightarrow\beta^{\frac{1}{2}}X^i$,
$\Phi_\rho\rightarrow\beta^{\frac{1}{2}}\Phi_\rho$,
$A\rightarrow\beta^{-1} A$, $c\rightarrow\beta^{\frac{1}{2}}c$ and
${\bar c}\rightarrow \beta^{\frac{1}{2}}{\bar c}$.  The fermions do
not need rescaling.  The path integral measure is now temperature
independent and, when the mass is zero, the only temperature
dependence is $\beta^{3}$ for the bosonic potential and $\beta^{3/2}$
for the fermionic potential. If the mass term is included it enters as
$\frac{\beta m^a}{\beta^{\frac{3}{2}}}$ in the potential with the
overall scales of $\beta^3$ and $\beta^{3/2}$ in the bosonic and
fermionic contributions respectively. The temperature dependence of
the model is now explicit.

The principal observable of the model is the energy\footnote{We divide
  by $N^2$ so that $E$ remains finite in the large-$N$ limit.},
$E=\langle H\rangle/N^2$.  Once the temperature dependence of the
model has been made explicit, as described above, one can then simply
note that $N^2 E$ is minus the derivative of logarithm of the
partition function with respect to $\beta$, returning to the original
variables one readily sees that in the path integral formulation:
\begin{align}
 E 
 &= \langle \varepsilon_b \rangle +  \langle \varepsilon_f \rangle \, ,\quad\hbox{where}
 \nonumber \\
 \varepsilon_b &=  \frac{3}{N\beta} 
 \int_0^\beta d\tau
 \bigg[
 \Tr \left( 
 -\frac{1}{4} [X^i,X^j]^2
 \right)
 \nonumber\\
 &\hspace{25mm}
 +\tr \left( 
   \bar{\Phi}^\rho X^{a\, 2} \Phi _\rho 
 - \bar{\Phi}^\rho [ \bar{X}^{\sigma \dot{\rho}}, X_{\rho \dot{\rho}}] \Phi_\sigma
 - \frac{1}{2}\bar{\Phi}^\rho \Phi_\sigma \bar{\Phi}^\sigma \Phi_\rho 
 + \bar{\Phi}^\rho \Phi_\rho \bar{\Phi}^\sigma \Phi_\sigma 
 \right) 
 \bigg]
 \, , 
 \nonumber \\
 \varepsilon_f &=  \frac{3}{2 N\beta} 
 \int_0^\beta d\tau
 \bigg[
 \Tr \left(
   \frac{1}{2} \lambda^{\dagger\rho} \gamma^a [X^a , \lambda_\rho] 
 - \frac{1}{2}\theta^{\dagger\dot{\rho}} \gamma^a [X^a , \theta_{\dot{\rho}}] 
 + \sqrt{2}i \varepsilon^{\rho \sigma} \theta^{\dagger\dot{\rho}} 
 [ X_{\sigma \dot{\rho}}, \lambda_\rho] 
 \right) 
 \nonumber\\
 &\hspace{25mm}
 +\tr \left( 
 -\chi^\dagger \gamma^a X^a \chi 
 - \sqrt{2}i \varepsilon^{\rho \sigma} \chi^\dagger \lambda_\rho \Phi_\sigma 
 - \sqrt{2}i \varepsilon_{\rho \sigma} 
 \bar{\Phi}^\rho \lambda^{\dagger\sigma} \chi 
 \right)
 \bigg]
 \, .
 \label{EnergyObservable}
\end{align}
We see only the potential contributes and the coefficients $3$ and
$3/2$ of the bosonic and fermionic terms arise from the differentiation.

As in \cite{Kawahara:2007ib},
there are two other interesting observables:
\begin{align}
 R^2 
 = \frac{1}{N\beta}\int_0^\beta d\tau \Tr X^{i\; 2}
 \, ,
 \qquad
 P
 = \frac{1}{N}\Tr \left( \exp \left[ i \beta A \right] \right)
 \, .
\end{align}
Here $R^2$ is a hermitian operator whose expectation value is a
measure of the extent of the eigenvalue distribution of the scalars $X^i$ and $P$ is
the Polyakov loop. Note: Path-ordering is not needed here for
the Polyakov loop as we consider $A$ in the static gauge.

Since the model has new degrees of freedom it is important to
consider other observables that capture properties of these
new fields. The natural candidates are
\begin{equation}
   r^2=\frac{1}{\beta N_f}\int_0^\beta d\tau \,
 \tr\bar\Phi^\rho\Phi_\rho\, ,
\label{defn_r2}
\end{equation}
which is the analogue of $R^2$ for the fundamental degrees of freedom,
and the condensate defined as
\begin{align}
 c^a(m)
 &=\frac{\partial}{\partial m^a}\left( -\frac{1}{N\beta}\log Z \right)
 =\left\langle\frac{1}{\beta}\int_0^\beta d\tau \tr \left\{
 2\bar\Phi^\rho (m^a-X^a)\Phi_\rho +\chi^\dagger \gamma^a \chi
 \right\}\right\rangle .
\end{align}
However, for us, with $m^a=0$, $c^a$ will be zero.
So our focus will be on the mass susceptibility
\begin{equation}
\langle{\mathcal C}^m\rangle:=\frac{\partial c^a}{\partial m^a} (0)\, ,
\label{mass_suseptibility_dfn}
\end{equation}
i.e.~the derivative with respect to $m^a$ with $a$ fixed (not summed over)
and evaluated at $m^a=0$ where
\begin{align}
{\mathcal C}^m
 &=
 \frac{2}{\beta}\int_0^\beta d\tau \tr \bar\Phi^\rho \Phi_\rho
 -\frac{N}{5\beta}\left(\int_0^\beta d\tau \tr \left\{
 -2\bar\Phi^\rho X^a\Phi_\rho +\chi^\dagger \gamma^a \chi
 \right\}\right)^2\, .
\label{condensate_susceptibility}
\end{align}
Here $a$ in \eqref{condensate_susceptibility} is summed over $a=1,\cdots,5$ 
and the same applies hereinafter.

\section{High-Temperature Expansion}
\label{High_Temp_Expans}

In this section, we develop the high-temperature expansion of the BD model.
For very high temperatures only the Matsubara zero modes,
i.e.~the zero modes in a Fourier expansion, survive and the model
reduces to a bosonic matrix model for these modes. A high-temperature
series expansion is therefore obtained by developing a perturbative
expansion of the model in the non-zero modes. The zero-modes must then
be treated non-perturbatively and this is done by Monte Carlo simulation.

Our strategy is therefore to expand the model in Matsubara modes, show
that the temperature can be seen as a coupling constant for these
modes and then integrate out the non-zero modes order by order in
perturbation theory to obtain an effective action for the zero modes,
which can then be treated non-perturbatively.

To obtain the series to second order we will only need one loop
computations.  The non-zero mode integration can be done analytically
and yields an effective action and observables in terms of the zero
temperature variables. As a final step the integration over these zero
modes must then be performed non-perturbatively via Monte Carlo
simulation.

The Fourier expansion of the fields is given by
\begin{align}
 &X^i(\tau)=\sum_{n\in \mathbb{Z}}X^i_{n}e^{2\pi i n\tau /\beta},
 \quad
 \lambda_{\rho}(\tau)=\sum_{r\in \mathbb{Z}+\frac{1}{2}}\lambda_{r\rho}e^{2\pi i r\tau /\beta},
 \quad
 \theta_{\dot\rho}(\tau)=\sum_{r\in \mathbb{Z}+\frac{1}{2}}\theta_{r\dot\rho}e^{2\pi i r\tau /\beta},
 \nonumber \\
 &c(\tau)=\sum_{n\in \mathbb{Z}, n\neq 0}c_{n}e^{2\pi i n\tau /\beta},
 \quad
 \bar c(\tau)=\sum_{n\in \mathbb{Z}, n\neq 0}\bar c_{n}e^{-2\pi i n\tau /\beta},
 \nonumber \\
 &\Phi_\rho(\tau)=\sum_{n\in\mathbb{Z}}\Phi_{n\rho}e^{2\pi i n\tau /\beta},
 \quad
 \chi(\tau)=\sum_{r\in \mathbb{Z}+\frac{1}{2}}\chi_{r}e^{2\pi i r\tau /\beta},
\end{align}
where thermal boundary conditions require that the bosons and ghosts are periodic in $\tau$ while the fermions are anti-periodic.

The action (\ref{action}) now takes the form of the sum of a zero mode action,
the kinetic term for the non-zero modes and an interaction term.
As discussed above, in the high-temperature limit, only the zero-modes
play a role.
As the temperature is lowered one can integrate out the non-zero modes
perturbatively with $\beta$ playing the role of a perturbation parameter. 
Using this procedure, the first two terms in the high-temperature expansion
of $E$, $\langle R^2\rangle$ and $\langle P\rangle$ for the BFSS model
were obtained in \citep{Kawahara:2007ib}. We follow the same method
here and obtain the corresponding expansion of these observables for
the BD model and for them the novel feature will be the additional
dependence on $N_f$, the number of
flavour multiplets. In addition we have the new observable
$\langle r^2 \rangle$ and $\langle{\mathcal C}^m\rangle$.

In order to develop the high-temperature series 
it is convenient to rescale the scalar fields in (\ref{action}) as follows
\begin{align}
 &X_0^i \rightarrow \beta^{-\frac{1}{4}}X_0^i \, ,\qquad
 A  \rightarrow \beta^{-\frac{1}{4}} A  \, ,\qquad 
 \Phi_0 \rightarrow \beta^{-\frac{1}{4}}\Phi_0 \, ,\qquad
 \nonumber\\ 
 &X_{n\neq 0}^i \rightarrow \beta^{\frac{1}{2}} X_{n\neq 0}^i \, , \qquad
 \Phi_{n\neq 0} \rightarrow \beta^{\frac{1}{2}} \Phi_{n\neq 0}  \, , \qquad
 c_n\rightarrow \beta^{\frac{1}{2}} c_n\, ,\qquad
 \bar c_n\rightarrow \beta^{\frac{1}{2}} \bar c_n\, ,
 \label{rescale}
\end{align}
while the fermions remain unchanged. This rescaling makes the coefficients of
the zero-mode terms and the kinetic terms independent of $\beta$ so
that one can concentrate on the $\beta$-dependence, which now appears
only in the interaction terms.

The rescaling (\ref{rescale}) can be understood as the rescaling
of section \ref{BD_model}, which was necessary to remove the
temperature dependence of the measure, 
followed by the further zero-mode rescaling
  $X_0^i\rightarrow\beta^{-\frac{3}{4}}X_0^i$,
  $\Phi_{0\rho}\rightarrow\beta^{-\frac{3}{4}}\Phi_{0\rho}$ and
  $A\rightarrow\beta^{\frac{3}{4}}A$. 
This zero-mode rescaling means the partition function becomes
\begin{equation}
  Z=\beta^{-\frac{3}{4}(8(N^2-1)+4N N_f)} \bar{Z}\, ,
\label{rescaled_Z}
\end{equation}
where ${\bar Z}$ is the partition function in terms of the rescaled
fields of (\ref{rescale}) and the only remaining temperature
dependence is in $S_\text{int}$. We can now develop the high-temperature series
by diagrammatic techniques with $\beta$ playing the role of a coupling.

The action is then written in terms of the variables of (\ref{rescale}), which
we will use for the remainder of the paper, as
\begin{equation}
  S=S_0+S_{\text{kin}}+S_{\text{int}}\; ,
\label{action_modes} 
\end{equation}
where $S_0$ is a zero-mode action
\begin{align}
 S_0 
 &= - \frac{N}{4} \Tr \left( [ X_0^i,X_0^j]^2 + 2[A,X_0^i]^2 \right) 
 + N\tr \bigg( \bar{\Phi}_0^\rho A^2 \Phi _{0\rho} 
 + \bar{\Phi}_0^\rho (X^a_0)^2\Phi_{0\rho} 
 \nonumber \\
 &\hspace{40mm}
 - \bar{\Phi}_0^\rho [ \bar{X}_0^{\sigma \dot{\rho}},X_{0\rho \dot{\rho}}]\Phi_{0\sigma} 
 - \frac{1}{2} \bar{\Phi}_0^\rho \Phi_{0\sigma} \bar{\Phi}_0^\sigma \Phi_{0\rho} 
 + \bar{\Phi}_0^\rho \Phi_{0\rho} \bar{\Phi}_0^\sigma \Phi_{0\sigma} \bigg) ,
 \label{action 0}
\end{align}
$S_{\text{kin}}$ is the kinetic part of the action for non-zero modes
\begin{align}
 S_{\text{kin}} = \sum_{n\neq 0} 
 &\frac{(2\pi n)^2N}{2} \bigg[ \Tr \left( 
 X_{-n}^a X_{n}^a
 +\bxo_{-n} \xon 
 +2\bar{c}_{-n} c_{n} \right) 
 +\tr \left( 2\bar{\Phi}_{-n}^\rho \Phi_{n\rho} \right) 
 \bigg]
 \nonumber\\
 &+ \sum_{r}\frac{2\pi irN}{2} \left[
 \Tr \left( \lambda_{-r}^{\dagger \rho} \lambda_{r\rho} 
 + \theta_{-r}^{\dagger \dot{\rho}} \theta_{r\dot{\rho}} \right) 
 + \tr \left( 2\chi_{-r} ^\dagger \chi_{r} \right) 
 \right] ,
 \label{action K}
\end{align}
and $S_{\text{int}}$ is the interaction part of the action. 
The terms quadratic in non-zero modes present in $S_{\text{int}}$ but not present in the BFSS model are
\begin{align}
 \iDelta S_{\text{int}}
 &= -N\beta^{\frac{3}{4}} (V_{1}^{(A)}+V_{1}^{(B)})
 -N\beta^{\frac{3}{2}} (V_{2}^{(A)}+V_{2}^{(B)}+V_{3})
 +O(\beta^{\frac{9}{4}}),
\label{GaussianPot}
\end{align}
where
\begin{align}
 &V_{1}^{(A)}
 =4\pi \sum_{n\neq 0} n \tr( \bar{\Phi}^\rho_{-n} A \Phi _{n\rho})
 +\sum_{r} \tr ( i\chi_{-r}^\dagger A \chi_{r} ) ,
 \nonumber \\
 &V_{1}^{(B)}
 =
 \sum_{r}
 \tr (
 \chi_{-r}^\dagger \gamma^a X_0^a \chi_r 
 +\sqrt{2}i \varepsilon^{\rho \sigma} \chi_{-r}^{\dagger} \lambda_{r\rho} \Phi_{0\sigma} 
 +\sqrt{2}i \varepsilon_{\rho \sigma} \bar{\Phi}_0^\rho \lambda^{\dagger\sigma}_{-r} \chi_{r} ) ,
 \nonumber\\
 &V_{2}^{(A)}
 =-\sum_{n\neq 0}
 \tr (\bar{\Phi}^\rho_{-n} A^2 \Phi_{n\rho}) ,
 \nonumber \\
 &V_{2}^{(B)}
 =-\sum_{n\neq 0}
 \tr \big( 
 \bar{\Phi}^\rho_{-n} X_0^{a\; 2} \Phi_{n\rho} 
 +\bar{\Phi}^\rho_{0} X^a_{-n} X^a_n \Phi_{0\rho}
 -\bar\Phi_{-n}^\rho 
 [\bar{X}_0^{\sigma \dot{\rho}},X_{0\rho \dot{\rho}}]\Phi_{n\sigma}
 -\bar\Phi_{0}^\rho 
 [\bar{X}_{-n}^{\sigma \dot{\rho}},X_{n\rho \dot{\rho}}]\Phi_{0\sigma}
 \nonumber\\
 &\hspace{28mm}
 -\bar{\Phi}_{-n}^\rho \Phi_{0\sigma} \bar{\Phi}_0^\sigma \Phi_{n\rho} 
 -\bar{\Phi}_{-n}^\rho \Phi_{n\sigma} \bar{\Phi}_0^\sigma \Phi_{0\rho} 
 +2\bar{\Phi}_{-n}^\rho \Phi_{0\rho} \bar{\Phi}_0^\sigma \Phi_{n\sigma} 
 +2\bar{\Phi}_{-n}^\rho \Phi_{n\rho} \bar{\Phi}_0^\sigma \Phi_{0\sigma} 
 \big) ,
 \nonumber\\
 &V_{3}
 =-\sum_{n\neq 0} \tr \Big[
 (
 \bar{\Phi}^\rho_{-n} X^a_n X^a_0 \Phi_{0\rho} 
 +\bar{\Phi}^\rho_{-n} X^a_{0} X^a_n \Phi_{0\rho} 
 +\bar{\Phi}^\rho_{0} X^a_{-n} X^a_0 \Phi_{n\rho} 
 +\bar{\Phi}^\rho_{0} X^a_{0} X^a_{-n} \Phi_{n\rho} 
 )
 \nonumber\\
 &\hspace{20mm}
 -(
 \bar\Phi_{-n}^\rho 
 [\bar{X}_n^{\sigma \dot{\rho}},X_{0\rho \dot{\rho}}]\Phi_{0\sigma}
 +\bar\Phi_{-n}^\rho 
 [\bar{X}_{0}^{\sigma \dot{\rho}},X_{n\rho \dot{\rho}}]\Phi_{0\sigma}
 \nonumber \\
 &\hspace{25mm}
 +\bar\Phi_{0}^\rho 
 [\bar{X}_{-n}^{\sigma \dot{\rho}},X_{0\rho \dot{\rho}}]\Phi_{n\sigma}
 +\bar\Phi_{0}^\rho 
 [\bar{X}_{0}^{\sigma \dot{\rho}},X_{-n\rho \dot{\rho}}]\Phi_{n\sigma}
 )
 \nonumber\\
 &\hspace{20mm}
 -\frac{1}{2}(
 \bar{\Phi}_{-n}^\rho \Phi_{0\sigma} \bar{\Phi}_n^\sigma \Phi_{0\rho} 
 +\bar{\Phi}_{0}^\rho \Phi_{-n\sigma} \bar{\Phi}_{0}^\sigma \Phi_{n\rho} 
 )
 +(
 \bar{\Phi}_{-n}^\rho \Phi_{0\rho} \bar{\Phi}_n^\sigma \Phi_{0\sigma} 
 +\bar{\Phi}_{0}^\rho \Phi_{-n\rho} \bar{\Phi}_{0}^\sigma \Phi_{n\sigma} 
 )
 \nonumber\\
 &\hspace{20mm}
 -\sum_{r} (
 \chi_{-r}^\dagger \gamma^{a} X_{-n}^a \chi_{r+n} 
 + \sqrt{2}i \varepsilon^{\rho \sigma} \chi_{-r}^{\dagger} \lambda_{r+n\, \rho} \Phi_{-n\sigma} 
 + \sqrt{2}i \varepsilon_{\rho \sigma} \bar{\Phi}_{-n}^\rho \lambda^{\dagger\sigma}_{-r} \chi_{r+n} )
 \Big] .
 \label{intV}
\end{align}
$V_3$ does not contribute to the expectation values of operators 
at next-leading order.  Two such vertices would be required and the
resultant contribution would therefore be of higher order in $\beta$.
Similarly, fermionic terms that involve only non-zero modes
also scale as $\beta^{\frac{3}{2}}$, and again contribute at a two and higher
loop order to the expectation values of observables.

The zero-mode action \eqref{action 0} corresponds to the bosonic part
of the original model (\ref{action}) dimensionally reduced to a point and plays an
important role in the ADHM construction as the solutions to $S_0=0$
with $\mathcal{D}^A=0$, where $\mathcal{D}^A$ is given in
\eqref{ADHM_D}, provide the ADHM data\cite{Tong:2005un}.
This zero-mode model is the flavoured bosonic version of the IKKT model\cite{Ishibashi:1996xs}.
We use the notation
$\langle\cdots\rangle_\text{\tiny DR}$ for the expectation value calculated with
this dimensionally reduced model. Thus for a generic observable, $O$ which is
a function of  $X_0^i$, $A$ and $\Phi_{0\rho}$ we denote
\begin{equation}
  \langle O \rangle_{_\text{\tiny DR}}=\frac{1}{\bar Z}\int [dX_0][d\bar\Phi_0][d \Phi_0] O {\rm e}^{-S_0}\, .
  \end{equation}
Furthermore we denote 
\begin{equation}
  \langle A B \rangle_\text{\tiny DR,c}=\langle A B \rangle_\text{\tiny DR}-\langle A \rangle_\text{\tiny DR}\langle B \rangle_\text{\tiny DR}\, ,
\end{equation}
and the subscript `c' denotes connected part.

Identities such as $\int \frac{ d }{dX_0}(X_0 O e^{-S_0})=0$
for $X_0$ and similar identities for $A$ and $\Phi_{0\rho}$ yield
the Ward-type identities:
\begin{eqnarray}
    \label{WardIdentities}
  9(N^2-1)\left\langle O\right\rangle_{{_\text{\tiny DR}}}- \left\langle\left. \lambda\frac{dS_0(\lambda X_0)}{d\lambda}\right\vert_{\lambda=1}
 O\right\rangle_\text{\tiny DR}+\left\langle\left. \lambda\frac{d O(\lambda X_0)}{d\lambda}\right\vert_{\lambda=1} \right\rangle_\text{\tiny DR}=0 \, ,
 \nonumber\\
    (N^2-1)\left\langle O\right\rangle_{_\text{\tiny DR}}- \left\langle\left. \lambda\frac{dS_0(\lambda A)}{d\lambda}\right\vert_{\lambda=1}
 O\right\rangle_{_\text{\tiny DR}}+\left\langle\left. \lambda\frac{d O(\lambda A)}{d\lambda}\right\vert_{\lambda=1} \right\rangle_{_\text{D\tiny R}} =0 \, ,
 \nonumber\\
   2N_f N\langle O\rangle_{_\text{\tiny DR}}-\left\langle\left. \lambda\frac{dS_0(\lambda \Phi_{0\rho})}{d\lambda}\right\vert_{\lambda=1}
 O\right\rangle_\text{\tiny DR}+\left\langle\left. \lambda\frac{d O(\lambda \Phi_{0\rho})}{d\lambda}\right\vert_{\lambda=1} \right\rangle_\text{\tiny DR} =0 \, .
  \end{eqnarray}
These identities can be used to simplify various expressions and in
particular one can see that one never needs to consider the
insertion of $S_0$ or
\begin{align}
 s_{0}
 &=N\Tr \Big( 
 -\frac{1}{4} [X_0^i,X_0^j]^2
 \Big) 
  +N\tr \Big( 
   \bar{\Phi}_0^\rho X_0^{a\, 2} \Phi _{0\rho }
 - \bar{\Phi}_0^\rho [ \bar{X}_0^{\sigma \dot{\rho}}, X_{0\rho \dot{\rho}}] \Phi_{0\sigma}
 \nonumber\\ 
&\hspace{60mm}
 - \frac{1}{2}\bar{\Phi}_0^\rho \Phi_{0\sigma} \bar{\Phi}_0^\sigma \Phi_{0\rho}
 + \bar{\Phi}_0^\rho \Phi_{0\rho} \bar{\Phi}_0^\sigma \Phi_{0\sigma} 
 \Big) 
\end{align}
with other correlators as they can be
eliminated by use of these identities. The simplest identities resulting from
(\ref{WardIdentities}) are that
\begin{equation}
 4 \langle S_0\rangle_\text{\tiny DR}=10(N^2-1)+4N_f N 
  \quad\hbox{and}\quad 
  4 \langle s_0\rangle_\text{\tiny DR}=8(N^2-1)+4N_fN \, ,
  \label{LeadingWardIdentities}
\end{equation}
the latter of which establishes the equivalent leading order expression
for the energy using (\ref{EnergyObservable}) as discussed in the next
section.

We will next present the leading high-temperature expansion of our observables.

\subsection{Leading order}
The expectation values of our observables to leading order are determined
solely by the zero modes. To this order the partition function is
given by (\ref{rescaled_Z}) with $\bar Z$ a constant.
We therefore have
\begin{align}
 E =\frac{3}{4}\beta^{-1}\left\{ 
 8\left( 1-\frac{1}{N^2} \right) + \frac{4N_f }{N} \right\}
 +O(\beta^{\frac{3}{2}})
 \, .
 \label{leading E}
\end{align}
The direct expression using (\ref{EnergyObservable}) gives $E=\frac{3}{N^2}\beta^{-1}\langle s_0 \rangle_\text{\tiny DR}+O(\beta^{3/2})$ and using
the second identity of (\ref{LeadingWardIdentities}) we see that this agrees with (\ref{leading E}).

Also, the leading terms in the $\beta$-expansion of $\langle R^2\rangle $ and
the expectation value of the Polyakov loop are 
\begin{align}
  \langle R^2 \rangle = \beta^{-\frac{1}{2}} 
 \left\langle \frac{1}{N}\Tr X_0^{i\; 2} \right\rangle_\text{\tiny DR}
 +O(\beta) 
 \, ,
 \quad
 \langle P \rangle
 = 1
 -\frac{\beta^{\frac{3}{2}}}{2}\left\langle \frac{1}{N}\Tr A^2 \right\rangle_\text{\tiny DR}
 +O(\beta^{3})
 \, .
\end{align}
For our new observables $\langle r^2\rangle$ and $\langle{\mathcal C}^m\rangle$
we have the leading contributions
\begin{equation}
 \langle r^2\rangle = \beta^{-\frac{1}{2}}\left\langle
 \frac{1}{N_f}\tr \bar\Phi_0^\rho \Phi_{0\, \rho}
 \right\rangle_\text{\tiny DR}+O(\beta)
\end{equation}
and
\begin{equation}
  \langle{\mathcal C}^m\rangle=2\beta^{-\frac{1}{2}}\left( 
 \left\langle \tr\bar\Phi_0^\rho \Phi_{0\rho} \right\rangle_{\text{\tiny DR}}
 -\frac{2N}{5}\left\langle (\tr\bar\Phi_{0}^\rho X_{0}^a\Phi_{0\rho})^2 \right\rangle_{\text{\tiny DR,c}}
 \right) +O(\beta)\, .
\end{equation}
Note: All the leading order contributions are purely bosonic,
since fermions decouple at high temperature. The necessary expectation
values are computed numerically via Monte Carlo simulation with the
action $S_0$ of equation (\ref{action 0}) and given in the tables 
in appendix \ref{AppendixB} for
different values of $N$ and $N_f$.

\subsection{Next-leading order}
The higher order contributions in the high-temperature expansion come
from integrating out the non-zero modes in (\ref{action_modes}).
The first subleading order is obtained by performing the gaussian integrals
over the non-zero modes, where the potential is truncated
as in (\ref{GaussianPot}), and expanding the resulting exponential and ratio of
determinants in terms of $\beta$.


Examining the action (\ref{action}) we see the fermionic terms can be written
in the form
\begin{align}
 &\int d\tau \bigg[
 \Tr\left(
 \frac{1}{2}\lambda^{\dag\rho}(D_\tau+\gamma^a{{\rm ad}\, X^a})\lambda_\rho
 +\frac{1}{2}\theta^{\dag\dot\rho}(D_\tau-\gamma^a {{\rm ad}\, X^a})\theta_{\dot\rho}
 \right)
 +\tr\left( \chi^\dag(D_\tau-\gamma^aX^a)\chi \right)
 \nonumber \\
 &\qquad\quad
 +\Tr\left( \theta^{\dag\dot\rho}J_{\dot\rho} \right)
 +\tr\left( \chi^\dag J+J^\dag\chi \right)
 \bigg] ,
\end{align}
and the commutator action of $X^a$ is denoted by `${\rm ad}\, X^a$'. 
Since $J_{\dot \rho}$ and $J$ are fermionic currents that
depend linearly on $\lambda_\sigma$, integrating
out $\theta_{\dot\rho}$ and $\chi$ gives the additional contributions
$-\frac{1}{2}\int J^{\dag\dot\rho}G_{\theta} J_{\dot\rho}$ and  $-\int J^\dag G_\chi J$
to the quadratic form for $\lambda_\rho$.
Here $G_{\lambda}$, $G_{\theta}$ and $G_{\chi}$ are Green's functions 
for $\lambda_\rho$, $\theta_{\dot\rho}$ and $\chi$, respectively.
These current-current terms will be of order $\beta^{3/2}$ and contribute
at the sub-leading order under consideration.

The non-zero modes can now be integrated out and to one loop we obtain
\begin{align}
 S_{eff}&=S_0 + \sum_{n\neq0}\frac{1}{2}\ln \frac{
 {\bf Det}[(1+\beta^{3/4}\frac{{\rm ad}\, A}{2\pi n})^2
 +\beta^{3/2}\frac{{\cal M}_X}{(2\pi n)^2}]
 {\bf Det}[(1+ \beta^{3/4} \frac{A}{2\pi n})^2 + \beta^{3/2} \frac{{\cal M}_{\phi}}{(2\pi n)^2}]}
 {{\bf Det}[1+\frac{{\rm ad}\, A}{2\pi n}]} \nonumber\\
 &\quad
 -\sum_{n=-\infty}^\infty\ln\bigg[
 {\bf Pf}\left\{ \epsilon\left(
 1+\beta^{3/4}\frac{-i\adj A +\gamma^a \adj X_0^a}{2\pi i (n+\frac{1}{2})}
 \right) \right\}
 \nonumber \\
 &\hspace{18mm}
 \times{\bf Pf}\left\{ \epsilon\left(
 1+\beta^{3/4}\frac{-i\adj A -\gamma^a \adj X_0^a}{2\pi i (n+\frac{1}{2})}
 \right) \right\}
 {\bf Det}\left( 
 1-\beta^{3/4}\frac{-i A +\gamma^a X_0^a}{2\pi i (n+\frac{1}{2})}
 \right)\bigg]
 \nonumber \\
 &\qquad
 +\sum_{n=-\infty}^{\infty}\beta^{3/2}\left(
 {\bf Tr}[G^n_\lambda \adj (X_0) G^{n}_\theta \adj (\bar X_0)]
 +2 {\bf Tr}[G^n_\lambda \Phi_0 G^{n}_\chi \bar\Phi_0]
 \right)\, ,
 \label{SeffLogDet_form}
\end{align}
with 
\begin{equation*}
 G^n_{\theta}=G^{n}_{\lambda} =G^{n}_\chi=\frac{1}{2\pi i(n+\frac{1}{2})}\, .
\end{equation*}
Equations
(\ref{GaussianPot}) with details in (\ref{intV}) specify the quadratic
forms whose determinants and Pfaffians enter in (\ref{SeffLogDet_form}), and 
here ${\bf Tr}$ is the operator trace.

In detail one has:
\begin{align}
 (\mathcal{M}_X X)^a_n 
 &= -\frac{1}{2}[X^b_0,[X^b_0,X^a_n]] +\frac{1}{2}[X^b_0,[X^a_0,X^b_n]] 
 +\frac{1}{2}[\bar X^{\rho\dot\rho}_0,[X_{0\rho\dot\rho}, X^a_{n}]] 
 -\Phi_{0\rho}\bar{\Phi}^\rho_{0}X^a_n,
 \nonumber \\
 (\mathcal{M}_X X)_{n \rho\dot\rho} 
 &= -\frac{1}{2}[X_{0\sigma\dot\rho},[\bar X_0^{\sigma\dot\sigma}, X_{n \rho\dot\sigma}]] 
 +\frac{1}{2}[X_{0\rho\dot\sigma},[\bar X_0^{\sigma\dot\sigma}, X_{n \sigma\dot\rho}]] 
 -\frac{1}{2}[X^a_0,[X^a_0, X_{n \rho\dot\rho}]] 
 \nonumber \\
 &\qquad
 -[\Phi_{0\rho}\bar{\Phi}^\sigma_{0}, X_{n \sigma\dot\rho}],
 \nonumber\\
 (\mathcal{M}_\Phi \Phi)_{n \rho}
 &=-X^{a\; 2}_0 \Phi_{n \rho} 
 +[\bar X_0^{\sigma\dot\rho}, X_{0\rho\dot\rho}] \Phi_{n \sigma} 
 \nonumber \\
 &\qquad
 +\Phi_{0\sigma} \bar\Phi^\sigma_0\Phi_{n\rho} +\Phi_{n\sigma} \bar\Phi_0^{\sigma}\Phi_{0\rho} 
 -2\Phi_{0\rho} \bar\Phi^\sigma_0\Phi_{n\sigma} -2\Phi_{m\rho} \bar\Phi_0^{\sigma}\Phi_{0\sigma} .
\end{align}
Expanding (\ref{SeffLogDet_form}) in $\beta$ we will obtain $S_{eff}$ with
\begin{equation}
 S_{eff}=S_0+\beta^{3/2}S_1\, .
\end{equation}
Let us look at the individual contributions in more detail.

After straightforward computations we find the contribution from the first
determinant in (\ref{SeffLogDet_form}) due to the integration over
non-zero $X^a$ modes gives
\begin{equation}
 S_1^X=\frac{N}{24}\left(16 \Tr X_0^{i\; 2}
 -18 \Tr A^2  \right) +\frac{5}{12} \frac{(N^2-1)}{N}\tr \bar{\Phi}_0^\rho \Phi_{0\rho}\, . 
\end{equation}
The ghost contribution similarly expanded gives
\begin{equation}
S_1^{g}=\frac{2N}{24} \Tr A^2\, ,
\end{equation}
and the contribution from the fundamental scalars is
\begin{equation}
S_1^{\Phi}=\frac{1}{12}\left(2N_f\Tr (X_0^{a\; 2} - A^2)+3 N \tr \bar{\Phi}_0^\rho \Phi_{0\rho} \right)\, .
\end{equation}
Putting the bosonic contributions together we have
\begin{equation}
S^{bos}_1=\frac{2N}{3}\left\{ 
 \Tr X_0^{i\; 2} -\Tr A^2
 + \left( 1-\frac{5}{8N^2} \right) \tr \bar{\Phi}_0^\rho \Phi_{0\rho}
 \right\}
 + \frac{N_f}{6}\left( 
 \Tr X_0^{a\; 2}
 - \Tr A^2  
 \right) \, .
\label{Sbos1}
\end{equation}

The fermionic contributions can similarly be evaluated to give
\begin{equation}
  S_1^{\theta}=S_1^{\lambda}=-N (\Tr X_0^{a\; 2} -\Tr A^2)
  \end{equation}
for the Pfaffian contribution to the integration
over $\theta_{\dot\rho}$ and $\lambda_\rho$. The $\theta$ current-current
contribution gives
\begin{equation}
  S_1^{J^{\dot\rho}G_\theta J_{\dot\rho}}=-2N \Tr(\bar X_0^{\rho \dot\rho} X_{0 \rho \dot\rho})\, .
\end{equation}
These three contributions together recombine to give an $SO(9)$ invariant term,
which is the fermionic part of $S_{1}^{ BFSS}$.

Next considering the $\chi$ determinant we find
\begin{equation}
S_1^{\chi}=-\frac{N_f}{2} (\Tr X_0^{a\;2}-\Tr A^2) \, ,
\end{equation}
and its current-current contribution is
\begin{equation}
S_1^{J^\dag G_\chi J}= -2\frac{(N^2-1)}{N}\tr \bar{\Phi}_0^\rho \Phi_{0\rho}\, .
\end{equation}
Putting all these fermionic contributions together we find
\begin{equation}
S_1^{fer}=-2N \left\{ 
 \Tr X_0^{i\; 2} -\Tr A^2
 + \left( 1-\frac{1}{N^2} \right) \tr \bar{\Phi}_0^\rho \Phi_{0\rho}
 \right\}
 - \frac{N_f}{2}\left( 
 \Tr X_0^{a\; 2}
 - \Tr A^2  
 \right) \, .
\label{Sfer1}
\end{equation}
Finally, defining ${\rm e}^{-\beta^{3/2} S_1}= 1+\beta^{3/2} {\mathcal{O}}$, so
for the supersymmetric model adding (\ref{Sbos1}) and (\ref{Sfer1})
we have
\begin{align}
\mathcal{O}
 =-\frac{4}{3}N\left\{ 
 \Tr X_0^{i\; 2} -\Tr A^2
 + \left( 1-\frac{19}{16N^2} \right) \tr \bar{\Phi}_0^\rho \Phi_{0\rho}
 \right\}
 + \frac{1}{3}N_f \left( 
 \Tr X_0^{a\; 2}
 - \Tr A^2  
 \right) \, .
 \label{O}
\end{align}
Similarly using (\ref{Sbos1}) we can define ${\mathcal{O}}_\text{bos}=-S_1^{bos}$, 
which we can write
\begin{equation}
 \mathcal{O}_\text{bos} = -\frac{1}{2} \mathcal{O}-\frac{3}{8N} \tr \bar{\Phi}_0^\rho \Phi_{0\rho}\, .
\end{equation}
These expressions will be useful in section \ref{Numerical_sims}. 

The partition function including the next to leading
corrections is now given by
\begin{equation}
  Z=\beta^{-\frac{3}{4}(8(N^2-1)+4N N_f)} (1+\beta^{3/2}\langle\mathcal{O}\rangle_\text{\tiny DR}) \bar{Z}\, ,
\label{rescaled_Z_with_sub}
\end{equation}
and the temperature dependence is explicit. We immediately have
the subleading correction to the energy $E$ by straightforward differentiation of (\ref{rescaled_Z_with_sub}) to obtain that
\begin{equation}
E =\frac{3}{4}\beta^{-1}\left\{ 
 8\left( 1-\frac{1}{N^2} \right) + \frac{4N_f }{N} \right\} -\frac{3}{2N^2}\beta^{1/2}\langle\mathcal{O}\rangle_\text{\tiny DR}\, . 
\end{equation}

Turning to the high-temperature behaviour of $R^2$ and the Polyakov loop,
the resulting expectation values are given by
\begin{align}
 &\langle R^2 \rangle
 = \beta^{-\frac{1}{2}}
 \left\langle \frac{1}{N}\Tr X_0^{i\; 2} \right\rangle_\text{\tiny DR}
 +\beta
 \left(
 \frac{3}{4}\left( 1-\frac{1}{N^2} \right)
 +\left\langle \frac{1}{N}(\Tr X_0^{i\; 2}) \mathcal{O} \right\rangle_{\text{\tiny DR,c}}
 \right)
 +O(\beta^{\frac{5}{2}}) \, ,
\label{R2}
\end{align}
and
\begin{align}
 \langle P \rangle
 &= 1-\beta^{\frac{3}{2}}\Bigg[
 \frac{1}{2}
 \left\langle \frac{1}{N}\Tr A^2 \right\rangle_\text{\tiny DR} 
 -\beta^{\frac{3}{2}}\left\{
 \frac{1}{4!}
 \left\langle \frac{1}{N}\Tr A^4\right\rangle_{\text{\tiny DR}} \!\!
 -\frac{1}{2}
 \left\langle \frac{1}{N}(\Tr A^2)\mathcal{O}\right\rangle_{\text{\tiny DR,c}} \!
 \right\}
 +O(\beta^{3})
 \Bigg] .
\end{align}
The constant $\frac{3}{4}(1-\frac{1}{N^2})$ is the contribution due to
the expectation value of the non-zero modes, which are traceless.

Our observable $\langle r^2\rangle$ is similarly given by
\begin{align}
 \langle r^2 \rangle
 =\beta^{-\frac{1}{2}} \left\langle
 \frac{1}{N_f}\tr \bar\Phi_0^\rho \Phi_{0\, \rho}
 \right\rangle _\text{\tiny DR}
 +\beta \left(
 \frac{1}{6}
 +\left\langle
 \frac{1}{N_f}\left(
 \tr \bar\Phi_0^\rho \Phi_{0\, \rho}
 \right)  \mathcal{O}
 \right\rangle _\text{\tiny DR,c}
 \right) 
 +O(\beta^{\frac{5}{2}}) \, ,
\end{align}
and its bosonic version is again obtained by replacing $\mathcal{O}$
with $\mathcal{O}_\text{bos}$.

In terms of Fourier modes we have
\begin{align}
 c^a(m)
 &=\left\langle\tr \bigg(
 2\beta^{-\frac{1}{2}}m^a\bar\Phi_0^\rho \Phi_{0\rho}
 +2\beta m^a\sum_{n\neq 0}\bar\Phi_{-n}^\rho \Phi_{n\rho}
 +\sum_{r}\chi_r^\dagger \gamma^a \chi_r
 -2\beta^{-\frac{3}{4}}\bar\Phi_{0}^\rho X_{0}^a\Phi_{0\rho}\right.
 \nonumber \\
 &\left. -2\beta^{\frac{3}{4}}\sum_{n}(
 \bar\Phi_{-n}^\rho X_{0}^a\Phi_{n\rho}
 +\bar\Phi_{0}^\rho X_{-n}^a\Phi_{n\rho}
 +\bar\Phi_{-n}^\rho X_{n}^a\Phi_{0\rho}
 )
 -2\beta^{\frac{3}{2}}\sum_{n,m}\bar\Phi_{-n}^\rho X_{n-m}^a\Phi_{m\rho}
 \bigg)\right\rangle .
\end{align}
However, we will restrict ourselves to the massless case and as discussed
$SO(5)$ invariance guarantees that this observable is zero
so we focus on the mass susceptibility, $\langle{\mathcal C}^m\rangle$.

Calculating ${\mathcal C}^m $
in the high-temperature expansion to the next to leading order
yields
\begin{align}
 \langle {\mathcal C}^m \rangle
 &=
 2\beta^{-\frac{1}{2}}\left( 
 \langle \tr\bar\Phi_0^\rho \Phi_{0\rho} \rangle_{\text{\tiny DR}}
 -\frac{2N}{5}\langle (\tr\bar\Phi_{0}^\rho X_{0}^a\Phi_{0\rho})^2 \rangle_{\text{\tiny DR,c}}
 \right)
 \nonumber \\
 &\qquad
 +2\beta
 \left(
 -\frac{N_f}{3}
 +\left\langle (\tr \bar\Phi_0^\rho\Phi_{0\rho}) \mathcal{O} \right\rangle_{\text{\tiny DR,c}}
 -\frac{2N}{5}\left\langle (\tr\bar\Phi_{0}^\rho X_{0}^a\Phi_{0\rho})^2
 \mathcal{O} \right\rangle_{\text{\tiny DR,c}}
 \right)
 +O(\beta^{\frac{5}{2}}).
 \label{condensate}
\end{align}
The contribution $-N_f/3$ in the second parentheses in \eqref{condensate}
contains both bosonic and fermionic contributions with the
fermionic contribution being $-N_f/2$,
while the bosonic contribution is $N_f/6$.
Therefore, the condensate susceptibility for the bosonic model 
is obtained from (\ref{condensate}) by replacing the numerical constant $-\frac{N_f}{3}$
by $\frac{N_f}{6}$ and $\mathcal{O}$  by $\mathcal{O}_\text{bos}$. The resulting expression is
\begin{align}
 \langle {\mathcal C}^{m} \rangle_\text{bos}
 &=2\beta^{-\frac{1}{2}}\left( 
 \langle \tr\bar\Phi_0^\rho \Phi_{0\rho} \rangle_{\text{\tiny DR}}
 -\frac{2N}{5}\langle (\tr\bar\Phi_{0}^\rho X_{0}^a\Phi_{0\rho})^2 \rangle_{\text{\tiny DR,c}}
 \right)
 \nonumber \\
 &\qquad
 +2\beta
 \left(
 \frac{N_f}{6}
 +\left\langle (\tr \bar\Phi_0^\rho\Phi_{0\rho}) \mathcal{O}_\text{bos} \right\rangle_{\text{\tiny DR,c}}
 -\frac{2N}{5}\left\langle (\tr\bar\Phi_{0}^\rho X_{0}^a\Phi_{0\rho})^2
 \mathcal{O}_\text{bos} \right\rangle_{\text{\tiny DR,c}}
 \right)
 +O(\beta^{\frac{5}{2}})\, .
\end{align}

An alternative to the above treatment is to work directly
with perturbation theory in $\beta$, but we believe the structure of
the computations is simpler in the above treatment. The contributions to 
$\mathcal{O}$, $E$, $R^2$ and $P$  from the pure BFSS model
were derived in \cite{Kawahara:2007ib} and when $N_f$ and
the fundamental fields are set to zero our results reproduce
the results reported there.

\section{High-temperature coefficients from numerical simulations}
\label{Numerical_sims}
In this section we express the coefficients, $\Xi_i$,
used in the high-temperature expansion of observables (see equations \eqref{def of obs} and \eqref{def of obs_bos}),
in terms of the primitive observables $\omega_{i}$, $\omega_{i,j}$ and $\omega_{i,j,k}$ ($i,j,k=1,...,6$) defined in equations \eqref{omegas} below.
We have the following expressions:
\begin{align}
 \Xi_1 
 &= -8(\omega_1 + \omega_3) 
 -\frac{2N_f}{N} \left\{ 
 \omega_1 + \left( 1-\frac{19}{16N^2}\right) \omega_4 
 \right\} , \nn \\
 \Xi_2 &= 5 \omega_1 + 4 \omega_3, \nn \\
 \Xi_3 &= \frac{3}{4}\left( 1-\frac{1}{N^2} \right) 
 +\frac{4 + \frac{N_f}{N}}{3}\left( 
 5 \omega_{1,1} + 15 \omega_{1,2} + 16 \omega_{1,3} 
 \right) 
 +\frac{16}{3} \left( 5 \omega_{1,3} + 4 \omega_{3,3} \right) \nn \\
 &\qquad
 +\frac{4N_f}{3N}\left( 1 - \frac{19}{16N^2} \right) 
 \left( 5 \omega_{1,4} + 4 \omega_{3,4} \right) , \nn \\
 \Xi_4 &= \frac{1}{2} \omega_1, \nn \\
 \Xi_5 &= \frac{1}{4!} \omega_6 
 -\frac{1}{2}\left\{ 
 \frac{4+\frac{N_f}{N}}{3} \left( 5 \omega_{1,2} - \omega_{1,1} \right) 
 +\frac{16}{3} \omega_{1,3} + 
 \frac{4 N_f}{3 N} \left( 1 - \frac{19}{16N^2} \right) \omega_{1,4}  
 \right\} , \nn \\
 \Xi_6 &=2 N_f \left(
 \omega_4 - 2\omega_{5,5}
 \right) , \nn \\
 \Xi_7 &= -\frac{2N_f}{3} 
 +8N_f \left\{ 
 \frac{4+\frac{N_f}{N}}{3} \omega_{1,4} 
 +\frac{4}{3} \omega_{3,4} 
 +\frac{1}{3} \left( 1-\frac{19}{16N^2} \right) \omega_{4,4} 
 \right\} , \nn \\
 \Xi_8 &= - 4 N_f \left\{ 
 \frac{4+\frac{N_f}{N}}{3} \left( \omega_{2,5,5} + 3 \omega_{1,5,5} \right) + \frac{16}{3} \omega_{3,5,5} 
 + \frac{4}{3}\left( 1-\frac{19}{16N^2} \right) \omega_{4,5,5} 
 \right\} , \nn \\
 \Xi_9 &= \omega_4 , \nn \\
 \Xi_{10} &= \frac{1}{4!}\omega_6 ,
 \nonumber \\
 \Xi_1^\text{bos} 
 &= -\frac{1}{2}\Xi_1-\frac{3N_f}{8N^3}\omega_4, \nn \\
 \Xi_3^\text{bos}
 &=\frac{3}{4}\left( 1-\frac{1}{N^2} \right) 
 -\frac{4 + \frac{N_f}{N}}{6}\left( 
 5 \omega_{1,1} + 15 \omega_{1,2} + 16 \omega_{1,3} 
 \right) 
 -\frac{8}{3} \left( 5 \omega_{1,3} + 4 \omega_{3,3} \right) \nn \\
 &\qquad
 -\frac{2N_f}{3N}\left( 1 - \frac{5}{8N^2} \right) 
 \left( 5 \omega_{1,4} + 4 \omega_{3,4} \right) 
 , \nn \\
 \Xi_5^\text{bos}
 &= \frac{1}{4!} \omega_6 
 +\frac{1}{4}\left\{ 
 \frac{4+\frac{N_f}{N}}{3} \left( 5 \omega_{1,2} - \omega_{1,1} \right) 
 +\frac{16}{3} \omega_{1,3} + 
 \frac{4 N_f}{3 N} \left( 1 - \frac{5}{8N^2} \right) \omega_{1,4}  
 \right\} , \nn \\
 \Xi_7^\text{bos}
 &= \frac{N_f}{3} 
 -4N_f \left\{ 
 \frac{4+\frac{N_f}{N}}{3} \omega_{1,4} 
 +\frac{4}{3} \omega_{3,4} 
 +\frac{1}{3} \left( 1-\frac{5}{8N^2} \right) \omega_{4,4} 
 \right\}, \nn \\
 \Xi_8^\text{bos}
 &=2N_f \left\{ 
 \frac{4+\frac{N_f}{N}}{3} \left( \omega_{2,5,5} + 3 \omega_{1,5,5} \right) + \frac{16}{3} \omega_{3,5,5} 
 + \frac{4}{3}\left( 1-\frac{5}{8N^2} \right) \omega_{4,5,5} 
 \right\} ,
 \label{def of Xi}
\end{align}
where
\begin{align}
 \omega_1 &= \frac{1}{N} \left\langle \Tr A^2 \right\rangle_\text{\tiny DR} , 
 \quad
 \omega_2 = \frac{1}{5N} \left\langle  \Tr (X_0^a)^2 \right\rangle_\text{\tiny DR} 
 =\omega_1,
 \quad
 \omega_3 = \frac{1}{4N} \left\langle \Tr \bar X_0^{\rho \dot\rho}X_{0\rho \dot\rho} \right\rangle_\text{\tiny DR} , 
 \nn \\
 \omega_4 &= \frac{1}{N_f} \left\langle   \tr \bar\Phi_0^\rho \Phi_{0\rho}   \right\rangle_\text{\tiny DR} ,
 \quad
 \omega_5 = \frac{1}{N} \left\langle  \tr \bar\Phi_0^\rho X^1_0 \Phi_{0\rho}   \right\rangle_\text{\tiny DR} =0, 
 \quad
 \omega_6 = \frac{1}{N} \left\langle  \Tr A^4   \right\rangle_\text{\tiny DR} ,
 \nn \\
 \omega_{1,1} &= \left\langle \left( \Tr A^2 \right)^2 \right\rangle_{\text{\tiny DR,c}} ,
 \quad
 \omega_{1,2} = \frac{1}{5} \left\langle  \Tr A^2  \Tr (X_0^a)^2   \right\rangle_{\text{\tiny DR,c}} ,
 \nonumber \\
 \qquad
 \omega_{1,3} &= \frac{1}{4} \left\langle  \Tr A^2  \Tr \bar X_0^{\rho \dot\rho}X_{0 \rho \dot\rho}   \right\rangle_{\text{\tiny DR,c}} ,
 \quad
 \omega_{1,4} =\frac{N}{N_f} \left\langle  \Tr A^2  \tr \bar\Phi_0^\rho \Phi_{0\rho}   \right\rangle_{\text{\tiny DR,c}} , \nn \\
 \omega_{3,3} &= \frac{1}{16} \left\langle 
 \left( \Tr \bar X_0^{\rho \dot\rho}X_{0 \rho \dot\rho} \right)^2
 \right\rangle_{\text{\tiny DR,c}},
 \quad
 \omega_{3,4} = \frac{N}{4N_f} \left\langle 
 \Tr \bar X_0^{\rho \dot\rho}X_{0 \rho \dot\rho} \tr \bar\Phi_0^\sigma \Phi_{0\sigma}
 \right\rangle_{\text{\tiny DR,c}} , \nn \\
 \omega_{4,4} &= \frac{N}{N_f} \left\langle  \left(  \tr \bar\Phi_0^\rho \Phi_{0\rho} \right)^2   \right\rangle_{\text{\tiny DR,c}} ,
 \quad
 \omega_{5,5} = \frac{N}{N_f} \left\langle \left( \tr \bar\Phi_0^\rho X^1_0 \Phi_{0\rho}\right)^2   \right\rangle_{\text{\tiny DR,c}} , \nn \\
 \omega_{1,5,5}&= \frac{N^2}{N_f}\left\langle \left( \Tr A^2 (\tr \bar\Phi_0^\rho X^1_0 \Phi_{0\rho}\right)^2   \right\rangle_{\text{\tiny DR,c}} ,
 \nonumber \\
 \omega_{2,5,5}&= \frac{N^2}{N_f}\left\langle \left( \Tr (X_0^1)^2 (\tr \bar\Phi_0^\rho X^1_0 \Phi_{0\rho}\right)^2   \right\rangle_{\text{\tiny DR,c}} ,\nn \\
 \omega_{3,5,5}&=  \frac{N^2}{4N_f}\left\langle 
 \Tr \bar X_0^{\rho \dot\rho}X_{0 \rho \dot\rho}
 \left( \tr \bar\Phi_0^\sigma X^1_0 \Phi_{0\sigma} \right)^2
 \right\rangle_{\text{\tiny DR,c}} ,
 \nonumber \\
 \omega_{4,5,5}&= \frac{N^2}{N_f}\left\langle 
 \left(  \tr \bar\Phi_0^\rho \Phi_{0\rho} (\tr \bar\Phi_0^\sigma X^1_0 \Phi_{0\sigma} \right)^2
 \right\rangle_{\text{\tiny DR,c}} .
 \label{omegas}
\end{align}

In terms of the $\Xi_i$ the observables of the full BD model (\ref{action})
become:
\begin{align}
 E
 &= \frac{3}{4}\beta^{-1}\left\{ 
 8\left( 1-\frac{1}{N^2} \right) + \frac{4N_f}{N}
 \right\}
 +\beta^{\frac{1}{2}}\Xi_1
 +O(\beta^{2}),
 \nonumber \\
 \langle R^2 \rangle
 &= \beta^{-\frac{1}{2}}
 \Xi_2
 +\beta \Xi_3
 +O(\beta^{\frac{5}{2}}) \, ,
 \nonumber \\
 \langle P \rangle
 &= 1-\beta^{\frac{3}{2}}\Bigg[
 \Xi_4
 -\beta^{\frac{3}{2}}\Xi_5
 +O(\beta^{3})
 \Bigg] ,
 \nonumber \\
 \langle r^2\rangle
 &= \beta^{-\frac{1}{2}} \Xi_9
 +\beta\left( \frac{1}{2}+\frac{\Xi_7}{2N_f} \right) 
 +O(\beta^{\frac{5}{2}}) \, ,
 \nonumber \\
 \langle {\mathcal C}^m\rangle &=
 \beta^{-\frac{1}{2}}\Xi_6
 +\beta (\Xi_7+\Xi_8)
 +O(\beta^{\frac{5}{2}})\, .
 \label{def of obs}
\end{align}
For the bosonic BD model we have
\begin{align}
 E_\text{bos}
 &= \frac{3}{4}\beta^{-1}\left\{ 
 8\left( 1-\frac{1}{N^2} \right) + \frac{4N_f}{N}
 \right\}
 +\beta^{\frac{1}{2}}\Xi_1^\text{bos}
 +O(\beta^{2}),
 \nonumber \\
 \langle R^2 \rangle_\text{bos}
 &= \beta^{-\frac{1}{2}}
 \Xi_2
 +\beta \Xi_3^\text{bos}
 +O(\beta^{\frac{5}{2}}) \, ,
 \nonumber \\
 \langle P \rangle_\text{bos}
 &= 1-\beta^{\frac{3}{2}}\Bigg[
 \Xi_4
 -\beta^{\frac{3}{2}}\Xi_5^\text{bos}
 +O(\beta^{3})
 \Bigg] ,
 \nonumber \\
 \langle r^2\rangle_\text{bos}
  &= \beta^{-\frac{1}{2}}
 \Xi_9
 +\beta\frac{1}{2N_f} \Xi_7^\text{bos}
 +O(\beta^{\frac{5}{2}})\, ,
 \nonumber\\
\langle {\mathcal C}^m\rangle_\text{bos}
 &=
 \beta^{-\frac{1}{2}}\Xi_6
 +\beta (\Xi_7^\text{bos}+\Xi_8^\text{bos})
 +O(\beta^{\frac{5}{2}}).
 \label{def of obs_bos}
\end{align}
The observables of interest for the high-temperature expansion
are all expressed in terms of $\Xi_i$ and $\Xi_{i}^\text{bos}$ listed above.
As discussed they are temperature independent and
depend only on $N$, the matrix dimension of the BFSS fields and
$N_f$, the number of flavour multiplets.
We computed their values for a range of $N$ and $N_f$ by hybrid Monte Carlo
simulation with the action $S_0$ given in \eqref{action 0}.
We tabulate our results for different $N$ and $N_f$.
We choose $N=4,6,8,9,10,12,14,16,18,20,32$ for $N_f=1$ 
and tabulate $\omega$'s, the building blocks of $\Xi_i$, 
in Table \ref{fig:tab1}.

From the results of Table \ref{fig:tab1} we extrapolate the $N$-dependence
of the $\omega$'s by fitting them with a function\footnote{Note that as expected
  we find it necessary to include a linear fall of in $1/N$ for large $N$.
  This is in contrast to the BFSS model, where the fall off is $1/N^2$},
$a+b/N+c/N^2$ (see Figure \ref{fig:a3} and \ref{fig:a7}).  
The limiting extrapolated values are included as the row $N=\infty$ in
Table \ref{fig:tab1}.

$\Xi_{i=1,\cdots,5,{\rm and 10}}$ naturally reduce to counterparts in the BFSS model
when the fundamental fields are removed. We extrapolate
$\Xi_{i=1,\cdots,5, {\rm and 10}}$ for $N_f=1,2,4,6,8,10,12,14,16$ for fixed
$N=12,14,16,18$
and $20$ to $N_f=0$ and find good agreement, to within the quoted errors,
with the measured values for their BFSS counterparts as quoted
in \cite{Kawahara:2007ib} .

Figure \ref{fig:b1} shows plots of the $\omega$'s against $N_f$ for each $N$
and we fit the dependence on $N_f$ basically with a quartic polynomial.
However, we find that higher order terms contribute
for some $\omega$'s and by using the fitting function
$a+b N_f+c e^{d N_f}$ we capture the dependence on $N_f$ over the range considered.

\section{The Bosonic Berkooz-Douglas model}
\label{BosonicBDmodel}
We are in the process of making a direct comparison of both 
the high-temperature regime of the BD model as determined by the above predictions and
the low-temperature regime as predicted by gauge/gravity with results from
a rational hybrid Monte Carlo simulation using the code
used in \cite{Filev:2015cmz}. We will present those results in a separate
paper as, apart from their value as a check on the code and the
computations presented here, they have additional physics that merits
a separate discussion.

For this paper we restrict our considerations to a comparison of the
results obtained here with those obtained from 
the bosonic Berkooz-Douglas model, 
whose Euclidean action is given by
\begin{align}
 S_\text{bos}=N\int_0^\beta d \tau  \, 
 &\Bigg[
 \Tr\left(\frac{1}{2}D_\tau X^a D_\tau X^a 
 + \frac{1}{2}D_\tau \bar{X}^{\rho \dot{\rho}} D_\tau X_{\rho \dot{\rho}} 
 -\frac{1}{4} [X^a,X^b]^2
 + \frac{1}{2}[X^a, \bxo][X^a,\xo]
 \right)
 \nonumber\\
 &\qquad +\tr \left( D_\tau \bar{\Phi}^\rho D_\tau \Phi_\rho
  + \bar{\Phi}^\rho (X^a-m^a)^2 \Phi _\rho \right)
 +\frac{1}{2}\Tr \sum_{A=1}^3\mathcal{D}^A\mathcal{D}^A
 \Bigg]\, .
 \label{BD_bos_action}
\end{align}

\begin{figure}[t]
\begin{center}
\includegraphics[width=0.35\paperwidth]{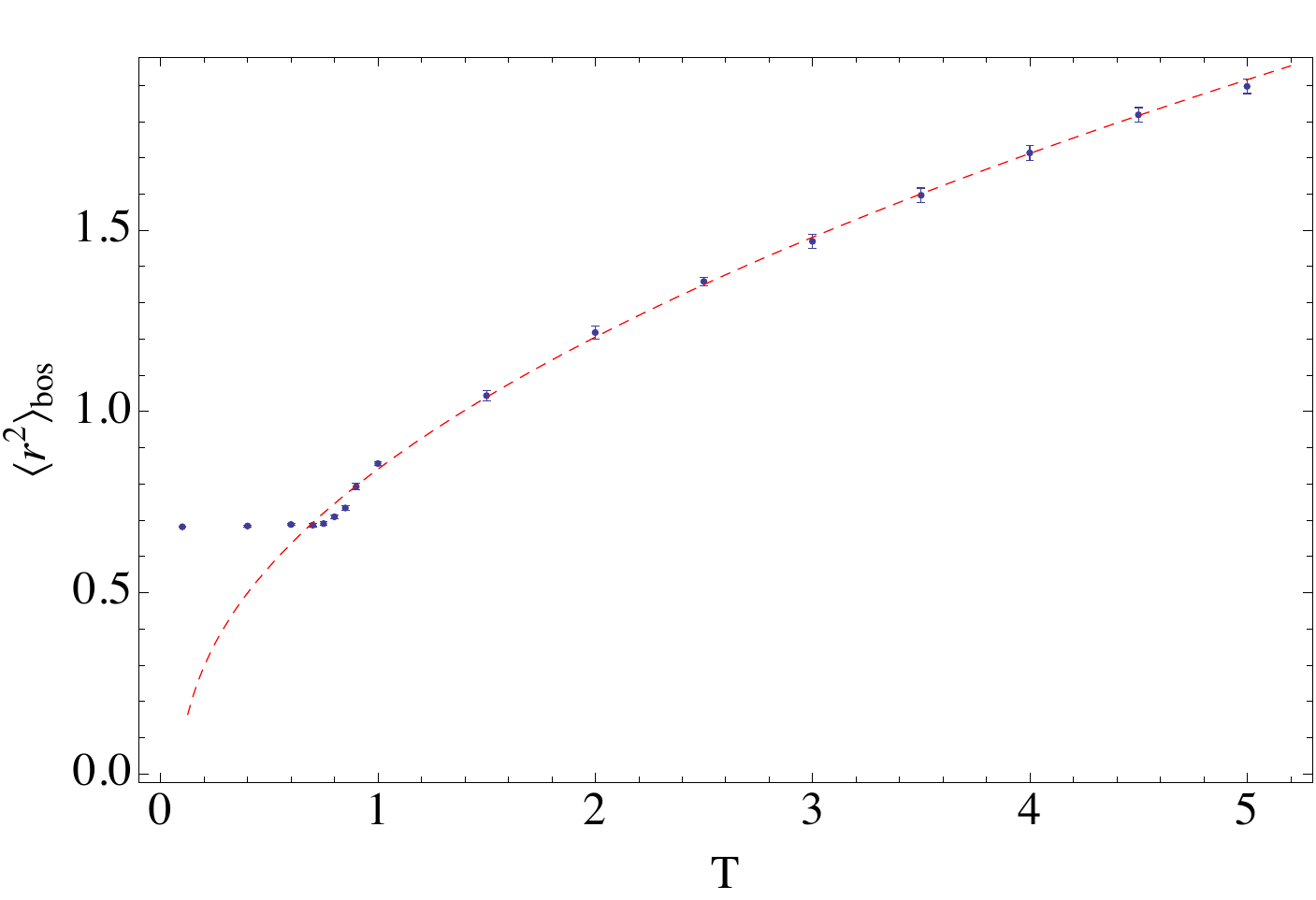}
\includegraphics[width=0.35\paperwidth]{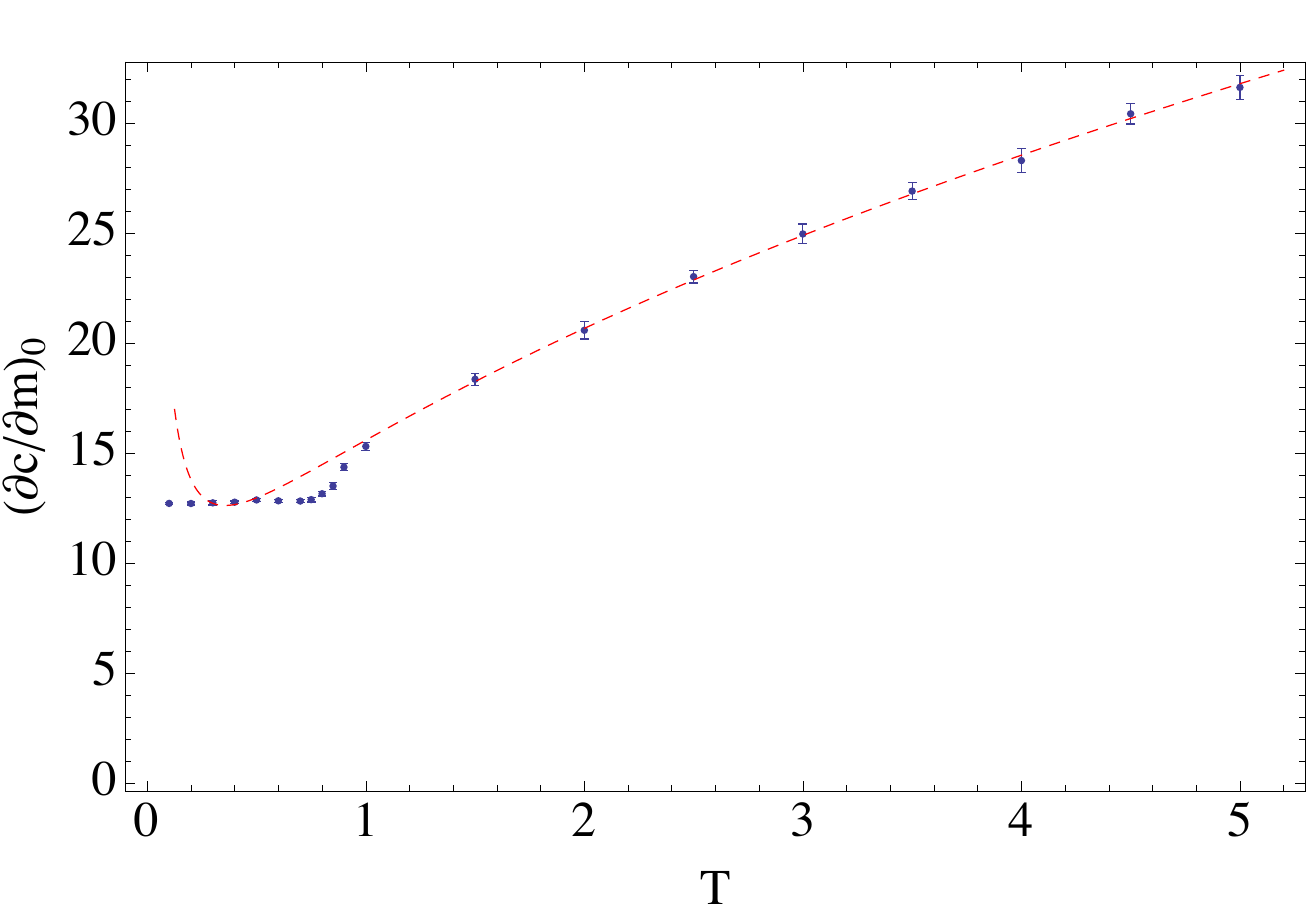}
\end{center}
\caption{Comparison of the high-temperature predictions for the
  fundamental observable $\langle r^2\rangle_\text{bos}$ and the derivative of
  the condensate at zero mass,
  $(\partial c/\partial m)_0=N\langle{\mathcal C}^m\rangle_\text{bos}$,
  with a Monte Carlo simulation of the bosonic BD model.
  The simulation is for $N_f=1$ and $N=10$.}
\label{fig:r2_and_dc_dm}
\end{figure}

Our comparison is presented in Figure~\ref{fig:r2_and_dc_dm}, where we
restrict our considerations to a high precision test with $N=10$ and
$N_f=1$.  As one can see from the figure the agreement is excellent.
Furthermore, the high $T$ expansion remains valid at temperatures as
low as $T\sim 1.0$. Below this temperature the figure shows evidence
of a phase transition. This is the phase transition of the
bosonic BFSS model.

From studies of the bosonic BFSS
model~\cite{Filev:2015hia,Kawahara:2007fn,Mandal:2009vz}
we know that it undergoes two phase transitions at
$T_{c2}=0.905\pm0.002$ and $T_{c1}=0.8761\pm 0.0003$. These are driven
by the gauge field $A$, which at high temperature behaves as one
of the $X^i$, while at low temperature it effectively disappears
from the system and can
be gauged away at zero temperature. As the temperature is increased
through $T_{c1}$ there is a deconfining phase transition with the
symmetry $A(t)\rightarrow A(t)+\alpha{\bf 1}$ broken and the
distribution of eigenvalues of the holonomy\footnote{The Polyakov loop,
  $P=\frac{1}{N}\Tr(U)$, where $U$ is the holonomy.} becomes non-uniform.
When the temperature reaches $T_{c2}$ the spectrum of the holonomy
becomes gapped and above this temperature the eigenvalues no longer
cover the entire $[0,2\pi]$ range. In the low-temperature phase the
bosonic BFSS model becomes a set of massive gaussian matrix
models with Euclidean action
\begin{equation}
 S_{eff}=N\int_{0}^{\beta}d\tau
  \Tr\left(\frac{1}{2}({\cal D}_{\tau}X^i)^2+m_A^2(X^i)^2\right)\, ,
  \label{Gaussian_Model}
\end{equation}
with the mass $m_A=1.965\pm0.007$.

For the flavoured model the BFSS transition is still present
and when the $X^i$ become massive they induce a mass for the
fundamental scalars and the induced bare mass for these
is estimated by integrating out the adjoint fields and expanding it to
quadratic order in $\Phi_\rho$. This gives a
mass $m_f^0=\sqrt{\frac{5}{2m_A}}\sim1.128$. However, the fundamental scalars
are still strongly interacting as they have a selfcoupling of order one
and we expect the bare mass to become significantly dressed.
We therefore estimate the physical mass of the scalars at
sufficiently low temperature by assuming that
they also can be described by a massive gaussian
with mass $m_f$, in which case 
\begin{equation}
\langle r^2\rangle_\text{bos}=\left\langle \frac{1}{\beta N_f}\int d\tau\, 
  \tr \bar\Phi^{\rho}\Phi_\rho \right\rangle_\text{bos}
  \simeq \frac{1}{m_f}\, .
\label{fundamental_r2}
\end{equation}
Note that the right-hand side of equation (\ref{fundamental_r2}) is independent of $\beta$ and from Figure \ref{fig:r2_and_dc_dm} we see that $\langle r^2\rangle_\text{bos}$ is more or less constant below the transition. A direct measurement
of the expectation value (\ref{fundamental_r2})
at $T=0.5$ gives $0.68618\simeq\frac{1}{m_f}$, which gives the
estimate $m_f\simeq1.4667$.

However, at zero temperature we can extract the masses for the different fields
by measuring their Green's function. To this end we set the holonomy to zero,
the parameter $\beta$ is now just the length of the time
circle and not an inverse temperature. 
\begin{figure}[t]
\begin{center}
\includegraphics[width=0.35\paperwidth]{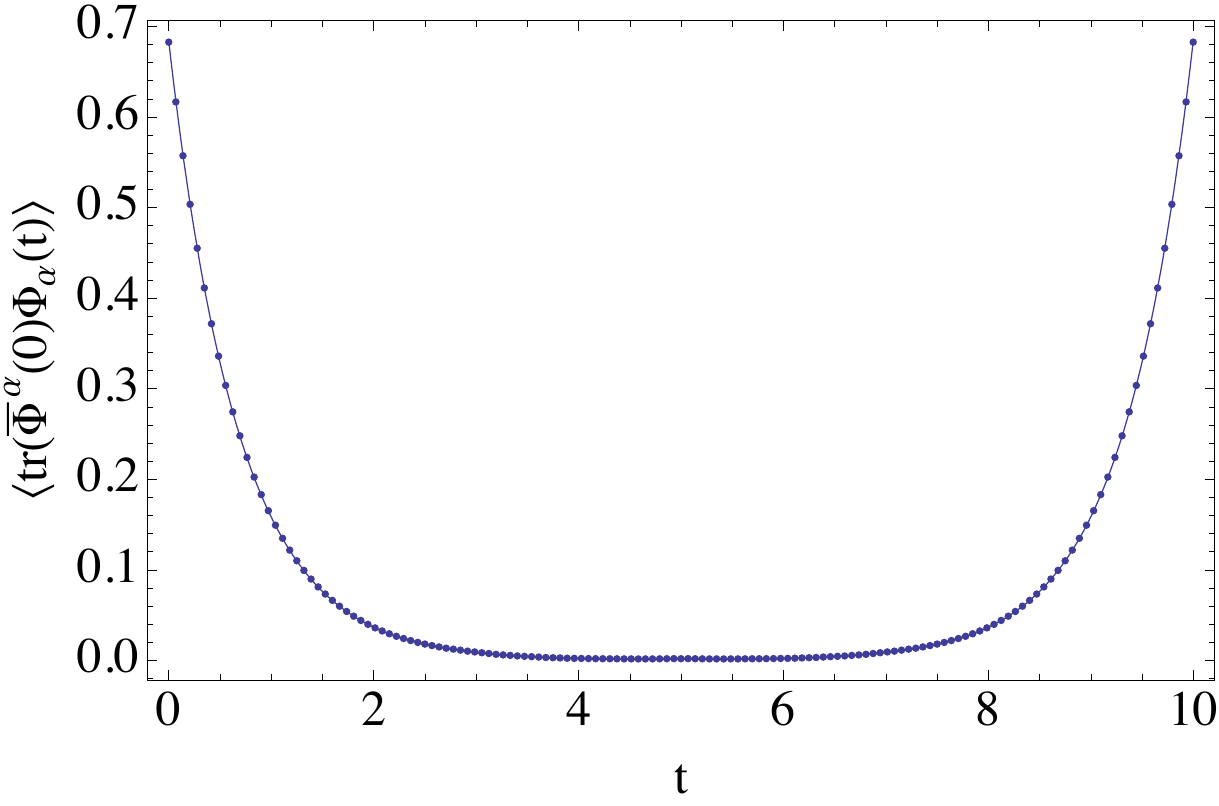}
\includegraphics[width=0.35\paperwidth]{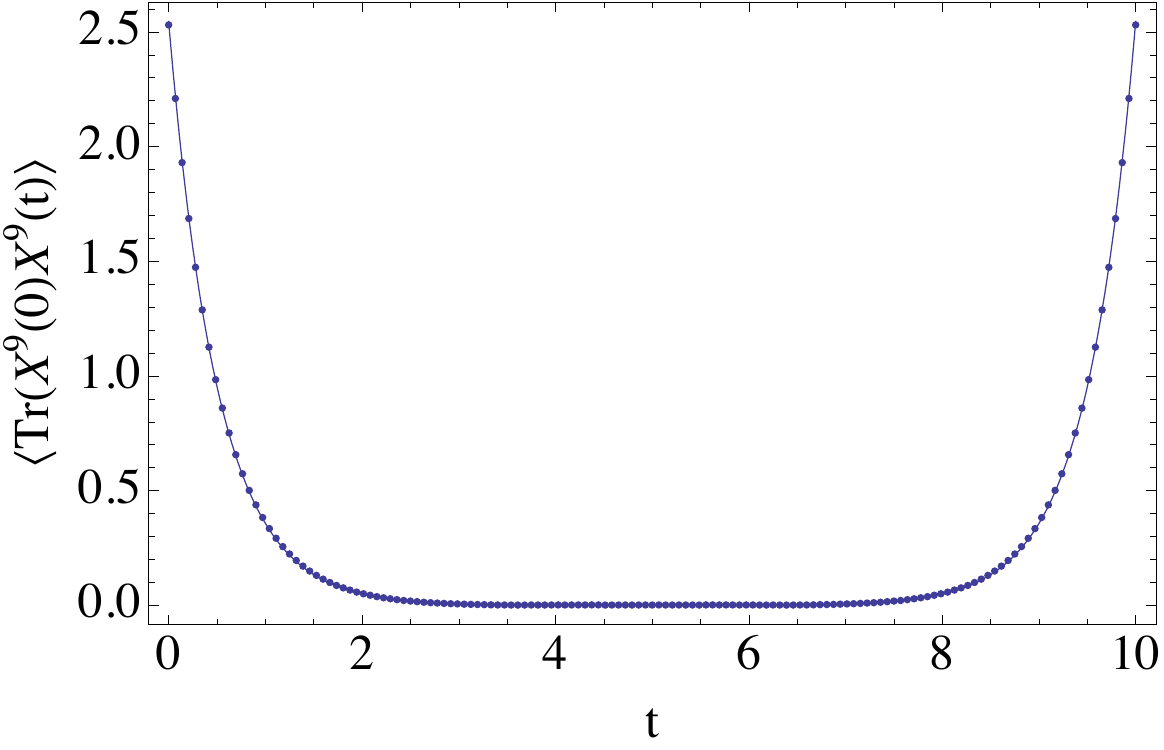}
\end{center}
\caption{Plots of the Green's functions equations (\ref{GreensPsi})
  and (\ref{GreensX}) for $\beta = 10$, $\Lambda =144$, $N =10$
  and $N_f = 1$. The fits correspond to $m_f = 1.461$ and $m^l_A = 2.001$.
}
\label{fig:greens}
\end{figure}
Because the $SO(9)$ symmetry of the bosonic BFSS model is broken down
to $SO(5)\times SO(4)$ there are now two adjoint masses, a longitudinal
mass, $m_A^l$, for the four $\xo$ and a transverse mass, $m_A^t$ , for the
five matrices $X^a$. 
In figure \ref{fig:greens} we present results for the Green's functions:
\begin{eqnarray}\label{GreensPsi}
\langle {\rm tr}\,\bar\Phi^{\rho}(0) \Phi_{\rho}(\tau)\rangle &=&\frac{N_f}{m_f}\, \frac{e^{-m_f\,\tau}+e^{-m_f(\beta-\tau)}}{1-e^{-\beta\,m_f}}\ ,\\
\langle {\rm Tr}X^9(0) X^9(\tau)\rangle &=&\frac{N}{2m^l_A} \,\frac{e^{-m^l_A\,\tau}+e^{-m^l_A(\beta-\tau)}}{1-e^{-\beta\,m^l_A}}\ ,
\label{GreensX}
\end{eqnarray}
where we have chosen the last of the four $SO(4)$ adjoint scalars. We have also
measured the longitudinal mass $m^l_A$ by measuring the correlator
for $X^1$ defined similarly to (\ref{GreensX}). 
The results for $\beta =10$, $\Lambda =144$, $N =10$ and $N_f = 1$
are  $m_A^l = 2.001 \pm 0.003$,  $m_A^t = 1.964 \pm 0.003$ and 
$m_f = 1.463 \pm 0.001$.  The prediction from assuming
that the adjoint fields are described by an action of the form (\ref{Gaussian_Model}) with different masses for the transverse and longitudinal matrices
is:
\begin{equation}
\langle R^2\rangle_\text{bos}\simeq\frac{5}{2 m^t_A}+\frac{4}{2m^l_A}=2.270 \pm 0.001\, ,
\end{equation}
which agrees well with the direct measurement where we find
$\langle R^2\rangle_\text{bos} = 2.261 \pm 0.005$.
Also the measured value of $m_f$  using (\ref{fundamental_r2}) predicts that
$\langle r^2\rangle_\text{bos} = 0.6836\pm0.0006$, 
which is in excellent agreement with
the measured value.

Note that this estimate of the mass $m_f$ is very close to the
one obtained from equation (\ref{fundamental_r2}). Also the slightly
different values of the adjoint masses $m^t_A$ and $m^l_A$ from the purely
BFSS case considered in equation (\ref{Gaussian_Model}) reflect the
presence of backreaction at $N_f/N = 0.1$. Observe also the closeness
of the transverse mass to the bosonic BFSS mass, which indicates
that the backreaction is strongest for the longitudinal
modes as one might expect.

We can now use this information to
estimate the value of $\langle{\mathcal C}^m\rangle_\text{bos}$ at zero
temperature.
Assuming that both $X^a$ and $\Phi_\rho$ are well approximated by massive
gaussians and using Wick's theorem on
\begin{align}
\mathcal{C}^m_\text{bos} =
 \frac{2}{\beta}\int_0^\beta d\tau \tr \bar\Phi^\rho \Phi_\rho
 -\frac{4N}{5\beta}\left(\int_0^\beta d\tau \tr \left\{\bar\Phi^\rho X^a\Phi_\rho 
 \right\}\right)^2
\label{condensate_susceptibility_bosonic}
\end{align}
to perform the contractions, we
obtain 
\begin{equation}
  \langle{\mathcal C}^m\rangle_\text{bos}\Big
  \vert_{T=0} 
 =\frac{2N_f}{m_f}-\frac{2N_f}{m_f^2m^t_A(2m_f+m^t_A)}=1.270\pm 0.001 \, .
\end{equation}
Finally, a direct measurement of the measured condensate shown in
Figure \ref{fig:r2_and_dc_dm} for $T=0.4,0.3,0.2$ and $T=0.1$ extrapolated
to $T=0$ gives $\langle{\mathcal C}^m\rangle_\text{bos}\big\vert_{T=0}= 1.268\pm0.003$, which is very
close to the predicted value and confirms the validity of the
gaussian approximation for both the adjoint and fundamental scalars.

\section{Conclusions}
\label{Conclusions}

We have obtained the first two terms in the high-temperature series
expansion for the Berkooz-Douglas model (BD model) for general adjoint
matrix size, $N$ and fundamental multiplet dimension, $N_f$. These
results should prove useful for future studies of this model. The
model is an ideal testing ground for many ideas of gauge/gravity
duality. The system is strongly coupled at low
temperature while at high temperature it is weakly coupled,
aside from the Matsubara zero-modes, which remain strongly coupled even
at high temperature. It is these modes that provide the residual
non-perturbative aspect of the current study. Their effect can be captured
in numerical coefficients that depend only on $N$ and $N_f$.

Once the coefficients are determined and tabulated (see appendix
\ref{AppendixB}) they can be used as input for the high-temperature
expansion of the observables of the BD model.  We have checked these
coefficients by comparing with a high precision simulation of the bosonic
version of the BD model which we simulated using the Hybrid Monte Carlo
approach. 
The coefficients in the high-temperature expansion of the bosonic model's observables are similarly determined by the tables presented in
appendix \ref{AppendixB}.
In fact the observable $\langle r^2\rangle_\text{bos}$
(see equation (\ref{defn_r2}))
and mass susceptibility (\ref{condensate_susceptibility_bosonic}) of the model,
shown in Figure \ref{fig:r2_and_dc_dm}, show that the agreement is excellent
even down to temperature one. Below this temperature the system
undergoes a set of phase transitions. These are essentially the two
phase transitions of the bosonic BFSS model.

We found that for $N_f/N=0.1$ our measurements were
sensitive to the backreaction of the fundamental fields on the
adjoint fields. This backreaction lifted the mass degeneracy of the transverse
and longitudinal adjoint fields. The transverse mass was essentially unaffected
by the backreaction being $m_A^t = 1.964 \pm 0.003$ while the longitudinal
mass was lifted to $m_A^l = 2.001 \pm 0.003$,

We found that using our understanding of the low-temperature phase of
the BFSS model as a system of massive gaussian quantum matrix models
we could predict the zero-temperature value of the mass
susceptibility (\ref{condensate_susceptibility_bosonic}).
The additional input that was required was the mass of the fundamental
fields which we found by direct measurement to be $m_f = 1.463 \pm 0.001$.

The zero-mode model used to obtain the high-temperature coefficients is
of independent interest as it is the potential that captures the ADHM data
in the theory of Yang-Mills instantons on the four-sphere, $S^4$.
It is the bosonic sector of the dimensional reduction of the BD model
to zero dimensions and is equivalent to a flavoured version of
the bosonic sector of the IKKT model. For this reason we refer to the
model as the flavoured bosonic IKKT model. The potential is always positive
semi-definite and the Higgs branch of its zero-locus is isomorphic to the
instanton moduli space \cite{Tong:2005un}.

There was some evidence for peculiar behaviour in the zero mode
model for $N_f \ge 2N $. We found that simulations required significant
fine tuning for $N_f\ge 2N$,  in that when using the same leapfrog step
length which gave $95\%$ acceptance rate for $N_f = 2N-1$ the acceptance rate
for $N_f \ge 2N$ fell to a fraction of this within a couple of
thousand sweeps and Ward identities we use as checks on the simulations
were not fulfilled. After tuning the simulation we found the
generated configurations had very long auto-correlation time. Also,
in fitting the dependence of the observables $\Xi_i$
on $N$ for a given $N_f$  we found evidence for a simple pole at $N=2N_f$.
Furthermore, one can see from the results tabulated in appendix
\ref{AppendixB} that they grow rapidly when the region $N_f=2N$ is approached.
We expect that these difficulties and the growth of observables as $N_f=2N$ is
approached are related to the singular structure of the instanton
moduli space, i.e.~the minimum of the
potential in  (\ref{action}) with $X^a=0$, ${\cal D}^A=0$.
We have not pursued this further in the current study as it would take us
too far afield, however, we believe it merits further attention.

Finally, our preliminary studies of the supersymmetric BD model
show \cite{in_prep} that, for some observables, the high-temperature
series expansion remains valid to lower temperatures than one
might expect. This validity of the high $T$ expansion at lower $T$
could provide alternative quasi-analytic estimates for observables
in the window where gauge/gravity duality is valid.

\section*{Acknowledgment}
D.O'C. thanks Stefano Kovacs and Charles Nash for helpful discussions on the ADHM construction. Support from Action MP1405 QSPACE of the COST foundation
is gratefully acknowledged.

\vfill\eject
\appendix

\section{Tables for the $\omega$'s. }
\label{AppendixB}
In this appendix we gather the numerical data from Monte Carlo simulations for different matrix sizes, $N$ 
and different numbers of flavour multiplets $N_f$ and present it in tabular form. 

\begin{table}[h]
\resizebox{\textwidth}{!}{
\begin{tabular}{|c|c|c|c|c|c|c|c|c|}
\hline 
  \multicolumn{9}{ |c| }{$N_f=1$} \\
  \hline
$N$ & $\omega_1$ & $\omega_3$ & $\omega_4$ & $\omega_6$ & $\omega_{1,1}$ & $\omega_{1,2}$ & $\omega_{1,3}$ & $\omega_{1,4}$ \\ 
\hline 
\hline
4 & \text{0.2201(1)} & \text{0.26221(6)} & \text{0.8974(2)} & \text{0.0972(1)} & \text{0.130(1)} & \text{-0.0055(4)} & \text{-0.0095(6)} & \text{-0.029(2)}\\ 
\hline 
6 & \text{0.23428(8)} & \text{0.26207(4)} & \text{0.8749(2)} & \text{0.11105(8)} & \text{0.140(1)} & \text{-0.0068(8)} & \text{-0.0097(8)} & \text{-0.029(3)}  \\ 
\hline 
8 & \text{0.24057(4)} & \text{0.26126(2)} & \text{0.8644(1)} & \text{0.11743(4)} & \text{0.146(1)} & \text{-0.0075(6)} & \text{-0.0097(6)} & \text{-0.029(3)} \\ 
\hline 
9 & \text{0.24246(3)} & \text{0.26087(2)} & \text{0.8606(1)} & \text{0.11938(4)} & \text{0.148(1)} & \text{-0.0078(7)} & \text{-0.0097(7)} & \text{-0.030(3)}  \\ 
\hline 
10 &  \text{0.243940(9)} & \text{0.260480(4)} & \text{0.85798(3)} & \text{0.120880(9)} & \text{0.1488(4)} & \text{-0.0079(2)} & \text{-0.0097(2)} & \text{-0.029(1)}  \\ 
\hline 
12 & \text{0.24608(2)} & \text{0.25987(1)} & \text{0.8539(1)} & \text{0.12309(3)} & \text{0.151(2)} & \text{-0.0082(9)} & \text{-0.0097(9)} & \text{-0.029(5)} \\ 
\hline 
14 & \text{0.24756(3)} & \text{0.25933(1)} & \text{0.8512(1)} & \text{0.12461(3)} & \text{0.152(3)} & \text{-0.008(1)} & \text{-0.01(1)} & \text{-0.029(6)}  \\ 
\hline 
16 & \text{0.24862(2)} & \text{0.258930(8)} & \text{0.8500(1)} & \text{0.12572(2)} & \text{0.154(2)} & \text{-0.009(1)} & \text{-0.01(1)} & \text{-0.029(7)}  \\ 
\hline 
18 & \text{0.24942(2)} & \text{0.258580(9)} & \text{0.84732(8)} & \text{0.12654(2)} & \text{0.154(3)} & \text{-0.009(1)} & \text{-0.01(2)} & \text{-0.030(8)}  \\ 
\hline 
20 &  \text{0.250070(4)} & \text{0.258300(2)} & \text{0.84604(2)} & \text{0.127220(5)} & \text{0.1553(9)} & \text{-0.0087(4)} & \text{-0.0097(5)} & \text{-0.029(3)}  \\ 
\hline 
32 & \text{0.252130(4)} & \text{0.257260(2)} & \text{0.84167(3)} & \text{0.129360(4)} & \text{0.158(2)} & \text{-0.0091(9)} & \text{-0.010(1)} & \text{-0.029(7)}   \\ 
\hline 
$\infty$ & \text{0.25530(3)} & \text{0.25533(3)} & \text{0.8346(1)} & \text{0.13282(7)} & \text{0.1619(2)} & \text{-0.00955(4)} & \text{-0.00959(3)} & \text{-0.0288(5)}  \\ 
\hline 
\end{tabular} 
}
\end{table}

\begin{table}[h]
\resizebox{\textwidth}{!}{
\begin{tabular}{|c|c|c|c|c|c|c|c|c|}
\hline 
  \multicolumn{9}{ |c| }{$N_f=1$} \\
  \hline
$N$ & $\omega_{3,3}$ & $\omega_{3,4}$ & $\omega_{4,4}$ & $\omega_{5,5}$ & $\omega_{1,5,5}$ & $\omega_{2,5,5}$ & $\omega_{3,5,5}$ & $\omega_{4,5,5}$ \\ 
\hline 
\hline
4 & \text{0.0403(6)} & \text{0.033(1)} & \text{0.305(2)} & \text{0.0779(1)} & \text{-0.0059(4)} & \text{0.033(1)} & \text{0.0020(5)} & \text{0.0483(6)}  \\ 
\hline 
6 & \text{0.0372(7)} & \text{0.031(2)} & \text{0.289(2)} & \text{0.07770(8)} & \text{-0.0064(8)} & \text{0.034(1)} & \text{0.0021(8)} & \text{0.0472(6)}  \\ 
\hline 
8 & \text{0.0361(8)} & \text{0.030(2)} & \text{0.280(2)} & \text{0.07726(9)} & \text{-0.006(1)} & \text{0.033(2)} & \text{0.002(1)} & \text{0.0461(8)} \\ 
\hline 
9 & \text{0.0357(7)} & \text{0.030(2)} & \text{0.277(2)} & \text{0.07710(9)} & \text{-0.007(2)} & \text{0.034(2)} & \text{0.002(2)} & \text{0.0458(9)} \\ 
\hline 
10& \text{0.0355(2)} & \text{0.030(1)} & \text{0.2752(6)} & \text{0.07699(3)} & \text{-0.0065(6)} & \text{0.0344(8)} & \text{0.0020(7)} & \text{0.0455(3)}\\ 
\hline 
12 & \text{0.0351(8)} & \text{0.029(4)} & \text{0.272(2)} & \text{0.07665(8)} & \text{-0.006(3)} & \text{0.031(3)} & \text{0.002(3)} & \text{0.0451(9)} \\ 
\hline 
14& \text{0.035(1)} & \text{0.029(5)} & \text{0.269(2)} & \text{0.07634(7)} & \text{-0.007(3)} & \text{0.035(4)} & \text{0.002(3)} & \text{0.0441(9)}  \\ 
\hline 
16& \text{0.035(1)} & \text{0.029(7)} & \text{0.267(3)} & \text{0.07629(8)} & \text{-0.007(5)} & \text{0.035(6)} & \text{0.002(5)} & \text{0.044(1)}  \\ 
\hline 
18 & \text{0.034(1)} & \text{0.029(8)} & \text{0.266(3)} & \text{0.07620(7)} & \text{-0.007(6)} & \text{0.038(7)} & \text{0.002(6)} & \text{0.044(1)}  \\ 
\hline 
20& \text{0.0343(4)} & \text{0.028(3)} & \text{0.2648(8)} & \text{0.07599(2)} & \text{-0.006(2)} & \text{0.034(2)} & \text{0.002(2)} & \text{0.0438(3)} \\ 
\hline 
32 & \text{0.0340(9)} & \text{0.028(7)} & \text{0.261(1)} & \text{0.07559(2)} & \text{-0.007(5)} & \text{0.035(5)} & \text{0.002(5)} & \text{0.0430(6)}  \\ 
\hline 
$\infty$ & \text{0.03352(6)} & \text{0.0272(1)} & \text{0.2542(1)} & \text{0.07479(6)} & \text{-0.0071(5)} & \text{0.036(2)} & \text{0.00188(6)} & \text{0.0416(2)}  \\ 
\hline 
\end{tabular}
}
\caption{Mean values of the $\omega$'s for $N_f=1$ were obtained from $3 \times 10^6$ Monte Carlo samples for all values of $N$ but $10 \ (3 \times 10^7 \mbox{samples})$ and $20,\ 32 \ (6 \times 10^7 \mbox{samples})$. Errors are estimated with the Jackknife resampling. $N=\infty$ values are the one extrapolated by quadratic polynomials: $a + b/N + c/N^2$. The quoted errors of this extrapolation are the fitting errors of the parameter $a$.}
\label{fig:tab1}
\end{table}

\newpage

In the remaining tables we tabulate fixed $N=9,12,14,16,18$ and $20$ while we vary $N_f$.  
Mean values of observable $\omega_{i}$, $\omega_{i,j}$ and $\omega_{i,j,k}$ ($i,j,k=1,...,6$) were obtained from $3 \times 10^6$ samples generated by hybrid Monte Carlo simulation of flavoured bosonic IKKT model with the action specified in \eqref{action 0}. Errors are estimated with the Jackknife resampling. 

\begin{table}[h]
\resizebox{\textwidth}{!}{
\begin{tabular}{|c|c|c|c|c|c|c|c|c|}
\hline 
  \multicolumn{9}{ |c| }{$N=9$} \\
  \hline
$N_f$ & $\omega_1$ & $\omega_3$ & $\omega_4$ & $\omega_6$ & $\omega_{1,1}$ & $\omega_{1,2}$ & $\omega_{1,3}$ & $\omega_{1,4}$ \\ 
\hline 
\hline
2 & \text{0.23205(3)} & \text{0.26942(2)} & \text{0.88939(9)} & \text{0.10936(3)} & \text{0.134(1)} & \text{-0.0062(7)} & \text{-0.0099(7)} & \text{-0.030(3)}  \\ 
\hline 
4&  \text{0.21144(2)} & \text{0.28898(2)} & \text{0.95582(8)} & \text{0.09091(2)} & \text{0.1093(9)} & \text{-0.0035(5)} & \text{-0.0107(6)} & \text{-0.033(2)} \\ 
\hline 
6 &  \text{0.19103(2)} & \text{0.31301(2)} & \text{1.03840(9)} & \text{0.07438(2)} & \text{0.0887(8)} & \text{-0.0015(4)} & \text{-0.0118(7)} & \text{-0.036(2)}  \\ 
\hline 
8 & \text{0.17041(2)} & \text{0.34381(3)} & \text{1.1462(1)} & \text{0.05941(2)} & \text{0.0709(6)} & \text{0.0002(4)} & \text{-0.0133(6)} & \text{-0.042(2)}  \\ 
\hline 
10 & \text{0.14909(2)} & \text{0.38552(5)} & \text{1.2945(2)} & \text{0.04570(1)} & \text{0.0551(6)} & \text{0.0016(4)} & \text{-0.0161(6)} & \text{-0.052(2)} \\ 
\hline 
12 &\text{0.12643(2)} & \text{0.44720(7)} & \text{1.5183(3)} & \text{0.03313(1)} & \text{0.0411(5)} & \text{0.0027(3)} & \text{-0.0208(6)} & \text{-0.070(2)}  \\ 
\hline 
14 &\text{0.10098(3)} & \text{0.5532(2)} & \text{1.9094(7)} & \text{0.02139(1)} & \text{0.0284(5)} & \text{0.0037(4)} & \text{-0.0311(6)} & \text{-0.108(3)}  \\ 
\hline 
16 & \text{0.06951(4)} & \text{0.8022(6)} & \text{2.838(2)} & \text{0.01036(1)} & \text{0.0165(5)} & \text{0.0045(4)} & \text{-0.0639(8)} & \text{-0.230(4)} \\  
\hline 
\end{tabular} 
}
\end{table}

\begin{table}[h]
\resizebox{\textwidth}{!}{
\begin{tabular}{|c|c|c|c|c|c|c|c|c|}
\hline 
  \multicolumn{9}{ |c| }{$N=9$} \\
  \hline
$N_f$ & $\omega_{3,3}$ & $\omega_{3,4}$ & $\omega_{4,4}$ & $\omega_{5,5}$ & $\omega_{1,5,5}$ & $\omega_{2,5,5}$ & $\omega_{3,5,5}$ & $\omega_{4,5,5}$ \\ 
\hline 
\hline
2 & \text{0.0384(7)} & \text{0.034(3)} & \text{0.310(3)} & \text{0.08089(7)} & \text{-0.006(1)} & \text{0.034(2)} & \text{0.002(2)} & \text{0.053(1)}  \\ 
\hline 
4& \text{0.0450(8)} & \text{0.044(3)} & \text{0.397(5)} & \text{0.08971(8)} & \text{-0.007(1)} & \text{0.033(2)} & \text{0.004(2)} & \text{0.072(3)}  \\ 
\hline 
6 & \text{0.054(1)} & \text{0.061(4)} & \text{0.54(1)} & \text{0.1015(1)} & \text{-0.007(2)} & \text{0.031(2)} & \text{0.006(3)} & \text{0.105(6)} \\ 
\hline 
8& \text{0.069(2)} & \text{0.091(6)} & \text{0.78(2)} & \text{0.1179(1)} & \text{-0.008(2)} & \text{0.032(3)} & \text{0.011(3)} & \text{0.16(1)}  \\ 
\hline 
10 & \text{0.095(3)} & \text{0.15(1)} & \text{1.25(6)} & \text{0.1423(2)} & \text{-0.010(2)} & \text{0.030(3)} & \text{0.019(5)} & \text{0.28(2)}  \\ 
\hline 
12 & \text{0.146(5)} & \text{0.27(2)} & \text{2.3(1)} & \text{0.1816(2)} & \text{-0.014(2)} & \text{0.030(3)} & \text{0.043(8)} & \text{0.55(4)}  \\ 
\hline 
14& \text{0.29(2)} & \text{0.66(6)} & \text{5.5(4)} & \text{0.2527(3)} & \text{-0.021(2)} & \text{0.027(3)} & \text{0.11(2)} & \text{1.3(1)} \\ 
\hline 
16 & \text{1.02(8)} & \text{2.9(3)} & \text{24.(2)} & \text{0.4275(9)} & \text{-0.043(4)} & \text{0.010(5)} & \text{0.52(7)} & \text{5.6(5)}  \\  
\hline
\end{tabular} 
}
\label{fig:tab9}
\end{table}

\begin{table}[h]
\resizebox{\textwidth}{!}{
\begin{tabular}{|c|c|c|c|c|c|c|c|c|}
\hline 
  \multicolumn{9}{ |c| }{$N=12$} \\
  \hline
$N_f$ & $\omega_1$ & $\omega_3$ & $\omega_4$ & $\omega_6$ & $\omega_{1,1}$ & $\omega_{1,2}$ & $\omega_{1,3}$ & $\omega_{1,4}$ \\ 
\hline 
\hline
2 &\text{0.23828(2)} & \text{0.26610(1)} & \text{0.87476(8)} & \text{0.11545(3)} & \text{0.141(2)} & \text{-0.0069(9)} & \text{-0.010(1)} & \text{-0.029(4)} \\ 
\hline 
4&  \text{0.22278(2)} & \text{0.27982(1)} & \text{0.92086(7)} & \text{0.10101(2)} & \text{0.121(1)} & \text{-0.0048(7)} & \text{-0.0103(8)} & \text{-0.031(4)}\\ 
\hline 
6 & \text{0.20742(2)} & \text{0.29562(2)} & \text{0.97439(8)} & \text{0.08770(2)} & \text{0.104(1)} & \text{-0.0030(7)} & \text{-0.0109(9)} & \text{-0.033(3)} \\ 
\hline 
8 &\text{0.19209(2)} & \text{0.31413(2)} & \text{1.0380(1)} & \text{0.07536(2)} & \text{0.089(1)} & \text{-0.0015(7)} & \text{-0.0118(9)} & \text{-0.037(3)} \\ 
\hline 
10 & \text{0.17660(2)} & \text{0.33637(3)} & \text{1.1149(1)} & \text{0.06387(2)} & \text{0.075(1)} & \text{-0.0002(6)} & \text{-0.0130(9)} & \text{-0.041(3)} \\ 
\hline 
12 & \text{0.16082(2)} & \text{0.36396(3)} & \text{1.2121(1)} & \text{0.05316(1)} & \text{0.0631(8)} & \text{0.0009(5)} & \text{-0.0146(8)} & \text{-0.046(3)}  \\ 
\hline 
14 &  \text{0.14454(2)} & \text{0.39950(4)} & \text{1.3391(2)} & \text{0.04314(1)} & \text{0.0518(7)} & \text{0.0019(5)} & \text{-0.0170(8)} & \text{-0.056(3)} \\ 
\hline 
16 & \text{0.12733(2)} & \text{0.44796(6)} & \text{1.5145(3)} & \text{0.03368(1)} & \text{0.0414(7)} & \text{0.0027(5)} & \text{-0.0208(8)} & \text{-0.070(4)}  \\  
\hline 
\end{tabular} 
}
\end{table}

\newpage

\begin{table}[h]
\resizebox{\textwidth}{!}{
\begin{tabular}{|c|c|c|c|c|c|c|c|c|}
\hline 
  \multicolumn{9}{ |c| }{$N=12$} \\
  \hline
$N_f$ & $\omega_{3,3}$ & $\omega_{3,4}$ & $\omega_{4,4}$ & $\omega_{5,5}$ & $\omega_{1,5,5}$ & $\omega_{2,5,5}$ & $\omega_{3,5,5}$ & $\omega_{4,5,5}$ \\ 
\hline 
\hline
2 & \text{0.0369(9)} & \text{0.032(4)} & \text{0.295(3)} & \text{0.07944(7)} & \text{-0.006(3)} & \text{0.033(3)} & \text{0.003(3)} & \text{0.050(2)}   \\ 
\hline 
4& \text{0.041(1)} & \text{0.039(4)} & \text{0.351(6)} & \text{0.08544(8)} & \text{-0.007(3)} & \text{0.033(3)} & \text{0.003(3)} & \text{0.062(4)}  \\ 
\hline 
6 & \text{0.047(1)} & \text{0.048(5)} & \text{0.43(1)} & \text{0.09269(9)} & \text{-0.006(3)} & \text{0.031(4)} & \text{0.004(4)} & \text{0.079(7)}  \\ 
\hline 
8 & \text{0.054(2)} & \text{0.061(7)} & \text{0.54(2)} & \text{0.10189(9)} & \text{-0.007(3)} & \text{0.031(3)} & \text{0.007(4)} & \text{0.11(1)}  \\ 
\hline 
10 & \text{0.065(3)} & \text{0.082(9)} & \text{0.70(3)} & \text{0.1136(1)} & \text{-0.008(3)} & \text{0.031(4)} & \text{0.010(6)} & \text{0.15(2)}  \\ 
\hline 
12 & \text{0.080(3)} & \text{0.11(1)} & \text{0.97(5)} & \text{0.1292(1)} & \text{-0.008(3)} & \text{0.028(3)} & \text{0.014(7)} & \text{0.21(2)} \\ 
\hline 
14 & \text{0.103(5)} & \text{0.17(2)} & \text{1.42(8)} & \text{0.1508(2)} & \text{-0.011(3)} & \text{0.033(4)} & \text{0.023(9)} & \text{0.32(4)}  \\ 
\hline 
16 & \text{0.144(7)} & \text{0.27(3)} & \text{2.3(1)} & \text{0.1816(2)} & \text{-0.014(3)} & \text{0.033(4)} & \text{0.04(1)} & \text{0.54(6)}  \\  
\hline
\end{tabular} 
}
\label{fig:tab12}
\end{table}

\begin{table}[h]
\resizebox{\textwidth}{!}{
\begin{tabular}{|c|c|c|c|c|c|c|c|c|}
\hline 
  \multicolumn{9}{ |c| }{$N=14$} \\
  \hline
$N_f$ & $\omega_1$ & $\omega_3$ & $\omega_4$ & $\omega_6$ & $\omega_{1,1}$ & $\omega_{1,2}$ & $\omega_{1,3}$ & $\omega_{1,4}$ \\ 
\hline 
\hline
2 & \text{0.24085(2)} & \text{0.26461(1)} & \text{0.86849(8)} & \text{0.11800(2)} & \text{0.143(2)} & \text{-0.007(1)} & \text{-0.010(1)} & \text{-0.029(5)} \\ 
\hline 
4&\text{0.22755(2)} & \text{0.27607(1)} & \text{0.90708(7)} & \text{0.10541(2)} & \text{0.127(2)} & \text{-0.0054(8)} & \text{-0.010(1)} & \text{-0.031(4)} \\ 
\hline 
6 &\text{0.21435(2)} & \text{0.28893(1)} & \text{0.95027(6)} & \text{0.09363(2)} & \text{0.111(2)} & \text{-0.0037(8)} & \text{-0.011(1)} & \text{-0.032(4)}   \\ 
\hline 
8 &\text{0.20119(2)} & \text{0.30355(1)} & \text{0.99992(7)} & \text{0.08262(2)} & \text{0.098(1)} & \text{-0.0023(8)} & \text{-0.011(1)} & \text{-0.035(4)}  \\ 
\hline 
10 & \text{0.18802(2)} & \text{0.32044(2)} & \text{1.05810(8)} & \text{0.07231(1)} & \text{0.085(1)} & \text{-0.0011(7)} & \text{-0.0121(9)} & \text{-0.038(4)}  \\ 
\hline 
12 & \text{0.17472(1)} & \text{0.34031(2)} & \text{1.12710(8)} & \text{0.06259(1)} & \text{0.0737(9)} & \text{0.0000(6)} & \text{-0.0132(9)} & \text{-0.041(3)} \\ 
\hline 
14 & \text{0.16118(1)} & \text{0.36434(2)} & \text{1.21160(9)} & \text{0.05344(1)} & \text{0.0632(9)} & \text{0.0010(6)} & \text{-0.0146(9)} & \text{-0.047(3)}  \\ 
\hline 
16 & \text{0.14723(1)} & \text{0.39416(3)} & \text{1.3179(1)} & \text{0.044771(8)} & \text{0.0536(8)} & \text{0.0018(5)} & \text{-0.0167(9)} & \text{-0.054(4)} \\  
\hline 
\end{tabular} 
}
\end{table}

\begin{table}[h]
\resizebox{\textwidth}{!}{
\begin{tabular}{|c|c|c|c|c|c|c|c|c|}
\hline 
  \multicolumn{9}{ |c| }{$N=14$} \\
  \hline
$N_f$ & $\omega_{3,3}$ & $\omega_{3,4}$ & $\omega_{4,4}$ & $\omega_{5,5}$ & $\omega_{1,5,5}$ & $\omega_{2,5,5}$ & $\omega_{3,5,5}$ & $\omega_{4,5,5}$ \\ 
\hline 
\hline
2 & \text{0.036(1)} & \text{0.031(5)} & \text{0.288(4)} & \text{0.07870(8)} & \text{-0.007(4)} & \text{0.035(4)} & \text{0.003(4)} & \text{0.049(2)}  \\ 
\hline 
4& \text{0.040(1)} & \text{0.037(5)} & \text{0.333(7)} & \text{0.08372(7)} & \text{-0.006(3)} & \text{0.032(4)} & \text{0.003(4)} & \text{0.058(4)}  \\ 
\hline 
6 & \text{0.044(2)} & \text{0.044(5)} & \text{0.39(1)} & \text{0.08958(9)} & \text{-0.006(4)} & \text{0.032(5)} & \text{0.004(5)} & \text{0.071(7)} \\ 
\hline 
8 & \text{0.050(2)} & \text{0.054(6)} & \text{0.47(2)} & \text{0.09655(8)} & \text{-0.008(3)} & \text{0.035(4)} & \text{0.005(5)} & \text{0.090(9)}  \\ 
\hline 
10 & \text{0.057(2)} & \text{0.067(8)} & \text{0.58(2)} & \text{0.1051(1)} & \text{-0.007(4)} & \text{0.030(4)} & \text{0.007(6)} & \text{0.12(2)}  \\ 
\hline 
12 & \text{0.066(2)} & \text{0.086(9)} & \text{0.73(3)} & \text{0.1156(1)} & \text{-0.007(4)} & \text{0.030(5)} & \text{0.010(8)} & \text{0.15(2)} \\ 
\hline 
14 & \text{0.080(3)} & \text{0.11(1)} & \text{0.97(4)} & \text{0.1297(1)} & \text{-0.009(4)} & \text{0.033(5)} & \text{0.015(8)} & \text{0.21(3)} \\ 
\hline 
16 & \text{0.099(5)} & \text{0.16(2)} & \text{1.34(8)} & \text{0.1473(1)} & \text{-0.010(4)} & \text{0.030(5)} & \text{0.02(1)} & \text{0.30(5)}  \\  
\hline
\end{tabular} 
}
\end{table}

\newpage

\begin{table}[h]
\resizebox{\textwidth}{!}{
\begin{tabular}{|c|c|c|c|c|c|c|c|c|}
\hline 
  \multicolumn{9}{ |c| }{$N=16$} \\
  \hline
$N_f$ & $\omega_1$ & $\omega_3$ & $\omega_4$ & $\omega_6$ & $\omega_{1,1}$ & $\omega_{1,2}$ & $\omega_{1,3}$ & $\omega_{1,4}$ \\ 
\hline 
\hline
2 & \text{0.24274(2)} & \text{0.263510(9)} & \text{0.86416(7)} & \text{0.11988(2)} & \text{0.146(2)} & \text{-0.008(1)} & \text{-0.010(1)} & \text{-0.029(6)}  \\ 
\hline 
4&\text{0.23108(2)} & \text{0.27333(1)} & \text{0.89680(5)} & \text{0.10870(2)} & \text{0.131(2)} & \text{-0.006(1)} & \text{-0.010(1)} & \text{-0.030(5)} \\ 
\hline 
6 & \text{0.21953(2)} & \text{0.28417(1)} & \text{0.93336(6)} & \text{0.09821(2)} & \text{0.117(2)} & \text{-0.004(1)} & \text{-0.010(1)} & \text{-0.031(4)}  \\ 
\hline 
8 &\text{0.20799(2)} & \text{0.29622(1)} & \text{0.97404(6)} & \text{0.08826(1)} & \text{0.104(2)} & \text{-0.0030(9)} & \text{-0.011(1)} & \text{-0.033(5)}  \\ 
\hline 
10 & \text{0.19650(1)} & \text{0.30983(1)} & \text{1.02060(6)} & \text{0.07890(1)} & \text{0.093(2)} & \text{-0.0018(8)} & \text{-0.012(1)} & \text{-0.036(4)} \\ 
\hline 
12 &\text{0.18493(1)} & \text{0.32538(2)} & \text{1.07400(7)} & \text{0.07002(1)} & \text{0.082(1)} & \text{-0.0008(8)} & \text{-0.012(1)} & \text{-0.038(5)}  \\ 
\hline 
14 & \text{0.17325(1)} & \text{0.34335(2)} & \text{1.13660(8)} & \text{0.06159(1)} & \text{0.072(1)} & \text{0.0002(8)} & \text{-0.013(1)} & \text{-0.042(4)} \\ 
\hline 
16 & \text{0.16138(1)} & \text{0.36458(2)} & \text{1.2110(1)} & \text{0.05360(1)} & \text{0.063(1)} & \text{0.0010(7)} & \text{-0.015(1)} & \text{-0.047(4)}  \\  
\hline 
\end{tabular} 
}
\end{table}

\begin{table}[h]
\resizebox{\textwidth}{!}{
\begin{tabular}{|c|c|c|c|c|c|c|c|c|}
\hline 
  \multicolumn{9}{ |c| }{$N=16$} \\
  \hline
$N_f$ & $\omega_{3,3}$ & $\omega_{3,4}$ & $\omega_{4,4}$ & $\omega_{5,5}$ & $\omega_{1,5,5}$ & $\omega_{2,5,5}$ & $\omega_{3,5,5}$ & $\omega_{4,5,5}$ \\ 
\hline 
\hline
2 & \text{0.036(1)} & \text{0.031(5)} & \text{0.283(4)} & \text{0.07823(7)} & \text{-0.007(5)} & \text{0.035(5)} & \text{0.002(5)} & \text{0.048(2)}  \\ 
\hline 
4 & \text{0.039(1)} & \text{0.035(5)} & \text{0.321(6)} & \text{0.08239(8)} & \text{-0.006(4)} & \text{0.032(6)} & \text{0.003(5)} & \text{0.056(5)}  \\ 
\hline 
6 & \text{0.043(2)} & \text{0.041(6)} & \text{0.37(1)} & \text{0.08736(8)} & \text{-0.007(5)} & \text{0.037(5)} & \text{0.003(6)} & \text{0.066(8)} \\ 
\hline 
8 & \text{0.047(2)} & \text{0.048(6)} & \text{0.43(1)} & \text{0.09295(8)} & \text{-0.007(4)} & \text{0.033(5)} & \text{0.005(6)} & \text{0.08(1)} \\ 
\hline 
10 & \text{0.052(2)} & \text{0.058(8)} & \text{0.51(2)} & \text{0.0996(1)} & \text{-0.008(5)} & \text{0.036(6)} & \text{0.006(8)} & \text{0.10(2)}  \\ 
\hline 
12 & \text{0.059(3)} & \text{0.07(1)} & \text{0.61(3)} & \text{0.1074(1)} & \text{-0.007(6)} & \text{0.033(6)} & \text{0.01(1)} & \text{0.12(2)} \\ 
\hline 
14 & \text{0.068(3)} & \text{0.09(1)} & \text{0.76(4)} & \text{0.1175(1)} & \text{-0.008(6)} & \text{0.031(7)} & \text{0.01(1)} & \text{0.16(3)} \\ 
\hline 
16 & \text{0.079(4)} & \text{0.11(2)} & \text{0.97(6)} & \text{0.1293(1)} & \text{-0.009(5)} & \text{0.034(6)} & \text{0.01(1)} & \text{0.21(4)} \\  
\hline
\end{tabular} 
}
\label{fig:tab16}
\end{table}

\begin{table}[h]
\resizebox{\textwidth}{!}{
\begin{tabular}{|c|c|c|c|c|c|c|c|c|}
\hline 
  \multicolumn{9}{ |c| }{$N=18$} \\
  \hline
$N_f$ & $\omega_1$ & $\omega_3$ & $\omega_4$ & $\omega_6$ & $\omega_{1,1}$ & $\omega_{1,2}$ & $\omega_{1,3}$ & $\omega_{1,4}$ \\ 
\hline 
\hline
2 & \text{0.24423(2)} & \text{0.262630(8)} & \text{0.86076(7)} & \text{0.12137(2)} & \text{0.147(3)} & \text{-0.008(1)} & \text{-0.010(2)} & \text{-0.029(7)} \\ 
\hline 
4&  \text{0.23383(2)} & \text{0.271220(9)} & \text{0.88931(5)} & \text{0.11132(2)} & \text{0.134(3)} & \text{-0.006(1)} & \text{-0.010(1)} & \text{-0.030(6)} \\ 
\hline 
6 & \text{0.22352(2)} & \text{0.280600(9)} & \text{0.92072(5)} & \text{0.10180(2)} & \text{0.121(2)} & \text{-0.005(1)} & \text{-0.010(1)} & \text{-0.032(5)} \\ 
\hline 
8 & \text{0.21329(1)} & \text{0.29087(1)} & \text{0.95536(6)} & \text{0.09279(1)} & \text{0.110(2)} & \text{-0.004(1)} & \text{-0.011(1)} & \text{-0.033(5)} \\ 
\hline 
10 &  \text{0.20303(1)} & \text{0.30226(1)} & \text{0.99401(6)} & \text{0.08418(1)} & \text{0.099(2)} & \text{-0.002(1)} & \text{-0.011(1)} & \text{-0.034(5)} \\ 
\hline 
12 & \text{0.19279(1)} & \text{0.31495(1)} & \text{1.03750(7)} & \text{0.07601(1)} & \text{0.089(2)} & \text{-0.0015(9)} & \text{-0.012(1)} & \text{-0.036(5)} \\ 
\hline 
14 & \text{0.18251(1)} & \text{0.32929(2)} & \text{1.08690(6)} & \text{0.06825(1)} & \text{0.080(2)} & \text{-0.0006(9)} & \text{-0.013(1)} & \text{-0.039(5)}  \\ 
\hline 
16 &  \text{0.17208(1)} & \text{0.34572(2)} & \text{1.14410(8)} & \text{0.060814(9)} & \text{0.071(1)} & \text{0.0002(8)} & \text{-0.013(1)} & \text{-0.042(5)} \\  
\hline 
\end{tabular} 
}
\end{table}

\newpage

\begin{table}[h]
\resizebox{\textwidth}{!}{
\begin{tabular}{|c|c|c|c|c|c|c|c|c|}
\hline 
  \multicolumn{9}{ |c| }{$N=18$} \\
  \hline
$N_f$ & $\omega_{3,3}$ & $\omega_{3,4}$ & $\omega_{4,4}$ & $\omega_{5,5}$ & $\omega_{1,5,5}$ & $\omega_{2,5,5}$ & $\omega_{3,5,5}$ & $\omega_{4,5,5}$ \\ 
\hline 
\hline
2 & \text{0.036(1)} & \text{0.030(7)} & \text{0.280(4)} & \text{0.07793(8)} & \text{-0.007(6)} & \text{0.037(7)} & \text{0.002(7)} & \text{0.047(3)}  \\ 
\hline 
4 & \text{0.038(2)} & \text{0.034(6)} & \text{0.312(7)} & \text{0.08155(7)} & \text{-0.006(6)} & \text{0.032(7)} & \text{0.003(7)} & \text{0.054(5)} \\ 
\hline 
6 & \text{0.041(2)} & \text{0.039(6)} & \text{0.35(1)} & \text{0.08566(8)} & \text{-0.007(6)} & \text{0.035(7)} & \text{0.003(7)} & \text{0.063(8)}  \\ 
\hline 
8  &\text{0.045(2)} & \text{0.045(8)} & \text{0.40(2)} & \text{0.09052(8)} & \text{-0.007(6)} & \text{0.033(6)} & \text{0.005(8)} & \text{0.07(1)}  \\ 
\hline 
10 & \text{0.049(2)} & \text{0.052(8)} & \text{0.46(2)} & \text{0.0959(1)} & \text{-0.006(7)} & \text{0.031(8)} & \text{0.00(1)} & \text{0.09(2)}  \\ 
\hline 
12 & \text{0.054(3)} & \text{0.06(1)} & \text{0.54(3)} & \text{0.10236(9)} & \text{-0.007(6)} & \text{0.031(7)} & \text{0.01(1)} & \text{0.11(2)} \\ 
\hline 
14 & \text{0.061(3)} & \text{0.07(1)} & \text{0.64(3)} & \text{0.1097(1)} & \text{-0.007(7)} & \text{0.030(8)} & \text{0.01(1)} & \text{0.13(3)} \\ 
\hline 
16 & \text{0.069(4)} & \text{0.09(1)} & \text{0.78(5)} & \text{0.1188(1)} & \text{-0.008(6)} & \text{0.032(7)} & \text{0.01(1)} & \text{0.16(4)}  \\  
\hline
\end{tabular} 
}
\label{fig:tab18}
\end{table}

\begin{table}[h]
\resizebox{\textwidth}{!}{
\begin{tabular}{|c|c|c|c|c|c|c|c|c|}
\hline 
  \multicolumn{9}{ |c| }{$N=20$} \\
  \hline
$N_f$ & $\omega_1$ & $\omega_3$ & $\omega_4$ & $\omega_6$ & $\omega_{1,1}$ & $\omega_{1,2}$ & $\omega_{1,3}$ & $\omega_{1,4}$ \\ 
\hline 
\hline
2 & \text{0.24535(2)} & \text{0.26193(1)} & \text{0.85819(9)} & \text{0.12249(2)} & \text{0.149(4)} & \text{-0.008(2)} & \text{-0.010(2)} & \text{-0.03(1)} \\ 
\hline 
4&  \text{0.23602(2)} & \text{0.26956(1)} & \text{0.88333(7)} & \text{0.11341(2)} & \text{0.136(3)} & \text{-0.006(2)} & \text{-0.010(2)} & \text{-0.031(8)} \\ 
\hline 
6 & \text{0.22673(2)} & \text{0.27779(1)} & \text{0.91089(6)} & \text{0.10474(2)} & \text{0.125(4)} & \text{-0.005(2)} & \text{-0.010(2)} & \text{-0.031(9)}  \\ 
\hline 
8 &  \text{0.21749(2)} & \text{0.28676(1)} & \text{0.94102(7)} & \text{0.09646(2)} & \text{0.114(3)} & \text{-0.004(2)} & \text{-0.011(2)} & \text{-0.032(8)} \\ 
\hline 
10 & \text{0.20828(2)} & \text{0.29653(1)} & \text{0.97400(6)} & \text{0.08855(1)} & \text{0.104(3)} & \text{-0.003(2)} & \text{-0.011(2)} & \text{-0.033(7)}   \\ 
\hline 
12 &  \text{0.19906(2)} & \text{0.30727(2)} & \text{1.01070(8)} & \text{0.08099(1)} & \text{0.095(3)} & \text{-0.002(1)} & \text{-0.011(2)} & \text{-0.036(9)} \\ 
\hline 
14 & \text{0.18984(2)} & \text{0.31919(2)} & \text{1.05150(9)} & \text{0.07377(1)} & \text{0.087(3)} & \text{-0.001(2)} & \text{-0.012(2)} & \text{-0.038(7)} \\ 
\hline 
16 &  \text{0.18053(2)} & \text{0.33253(2)} & \text{1.0975(1)} & \text{0.06683(1)} & \text{0.078(2)} & \text{0.000(2)} & \text{-0.013(2)} & \text{-0.039(8)} \\  
\hline 
\end{tabular} 
}
\end{table}

\begin{table}[h]
\resizebox{\textwidth}{!}{
\begin{tabular}{|c|c|c|c|c|c|c|c|c|}
\hline 
  \multicolumn{9}{ |c| }{$N=20$} \\
  \hline
$N_f$ & $\omega_{3,3}$ & $\omega_{3,4}$ & $\omega_{4,4}$ & $\omega_{5,5}$ & $\omega_{1,5,5}$ & $\omega_{2,5,5}$ & $\omega_{3,5,5}$ & $\omega_{4,5,5}$ \\ 
\hline 
\hline
2 & \text{0.035(2)} & \text{0.03(1)} & \text{0.278(6)} & \text{0.07754(8)} & \text{-0.006(8)} & \text{0.032(9)} & \text{0.002(8)} & \text{0.046(3)} \\ 
\hline 
4 & \text{0.038(2)} & \text{0.03(1)} & \text{0.305(9)} & \text{0.08075(7)} & \text{-0.007(7)} & \text{0.037(8)} & \text{0.003(8)} & \text{0.052(5)} \\ 
\hline 
6 & \text{0.040(3)} & \text{0.037(9)} & \text{0.34(1)} & \text{0.08440(8)} & \text{-0.007(7)} & \text{0.037(9)} & \text{0.002(9)} & \text{0.059(9)} \\ 
\hline 
8 & \text{0.043(3)} & \text{0.04(1)} & \text{0.38(2)} & \text{0.08853(8)} & \text{-0.007(7)} & \text{0.031(8)} & \text{0.004(9)} & \text{0.07(1)}  \\ 
\hline 
10 & \text{0.047(3)} & \text{0.05(1)} & \text{0.43(2)} & \text{0.0930(1)} & \text{-0.007(8)} & \text{0.03(1)} & \text{0.00(1)} & \text{0.08(2)} \\ 
\hline 
12 & \text{0.051(4)} & \text{0.06(2)} & \text{0.49(4)} & \text{0.0984(1)} & \text{-0.007(8)} & \text{0.03(1)} & \text{0.01(1)} & \text{0.09(3)}  \\ 
\hline 
14 & \text{0.056(5)} & \text{0.07(2)} & \text{0.57(5)} & \text{0.1044(1)} & \text{-0.008(9)} & \text{0.031(9)} & \text{0.01(1)} & \text{0.11(4)}  \\ 
\hline 
16 & \text{0.062(6)} & \text{0.08(2)} & \text{0.66(7)} & \text{0.1114(1)} & \text{-0.008(9)} & \text{0.03(1)} & \text{0.01(2)} & \text{0.14(4)} \\  
\hline
\end{tabular} 
}
\caption{The tables for $N=9$ to $N=20$ were obtained from Monte Carlo simulations with $3 \times 10^6$ samples and errors are estimated using Jackknife resampling.}
\label{fig:tab20}
\end{table}

\newpage

\begin{figure}[htbp]
\centering
  \makebox[\textwidth]{\includegraphics[width=.9\paperwidth]{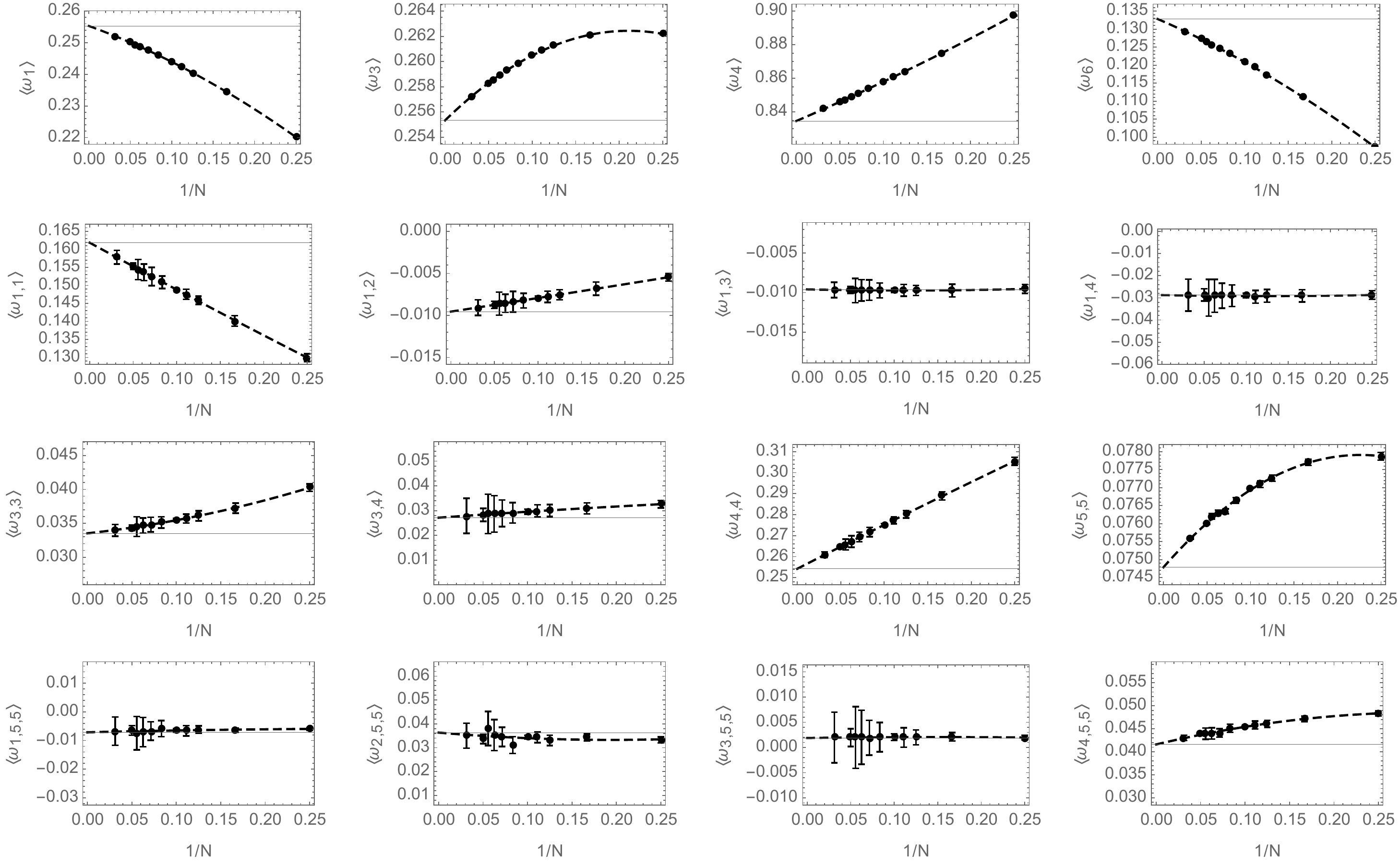}}
\caption{Mean values of the $\omega$'s plotted against $N$ with $N_f=1$. 
Dashed lines correspond to fits of the form $a + b/N + c/N^2$ while vertical lines correspond to $N \rightarrow \infty$ values obtained from those fits. Errors are estimated with the Jackknife resampling. Its values are quite small and it is determined rather precisely in the tables. }
\label{fig:a3}
\end{figure}

\newpage

\begin{figure}[htbp]
\centering
  \makebox[\textwidth]{\includegraphics[width=.9\paperwidth]{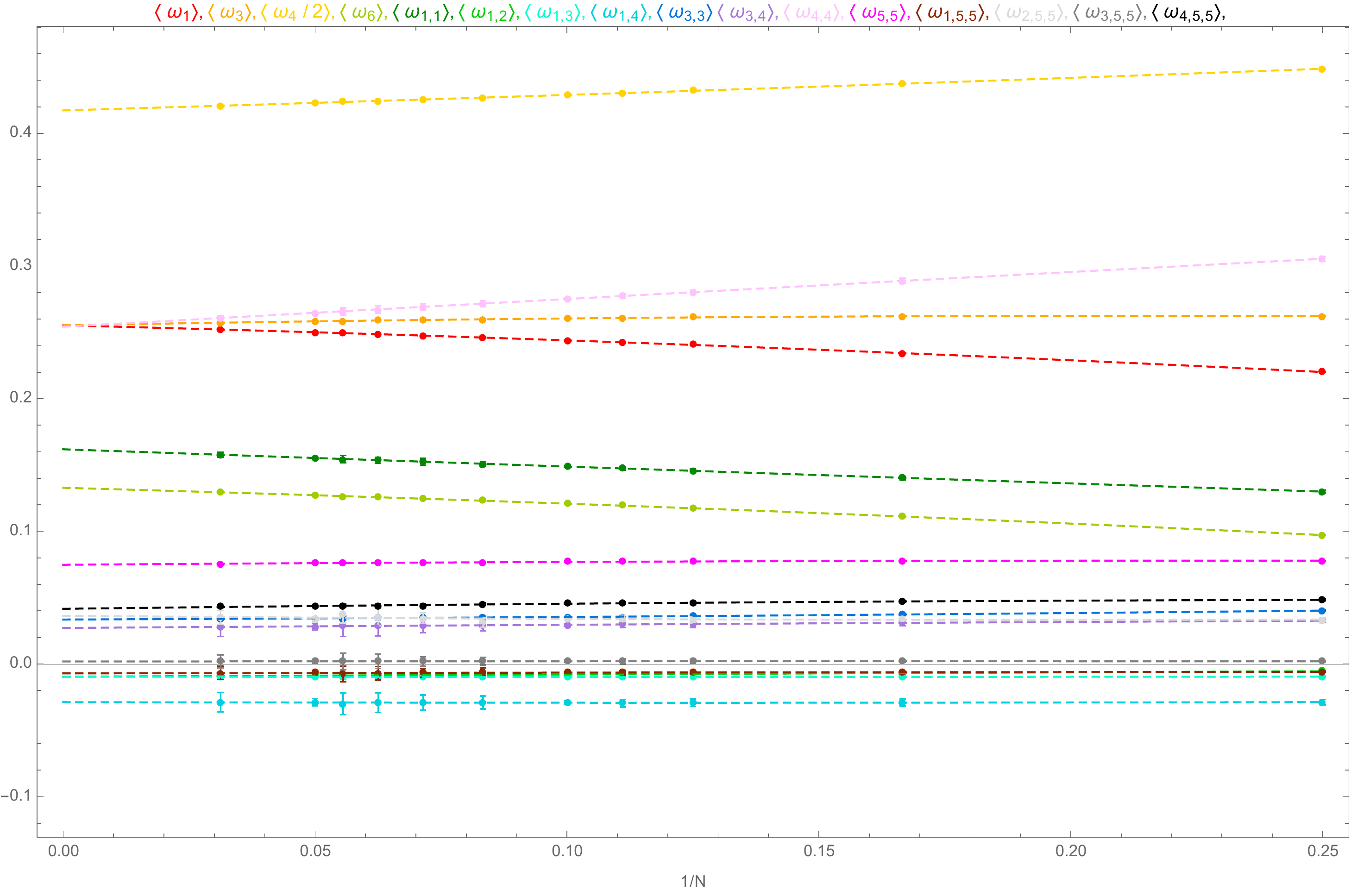}}
\caption{Mean values of the $\omega$'s plotted against $N$ with $N_f=1$. Dashed lines correspond to fits of the form $a + b/N$. Errors are estimated with the Jackknife resampling.}
\label{fig:a7}
\end{figure}

\newpage

\begin{figure}[htbp]
\centering
  \makebox[\textwidth]{\includegraphics[width=.9\paperwidth]{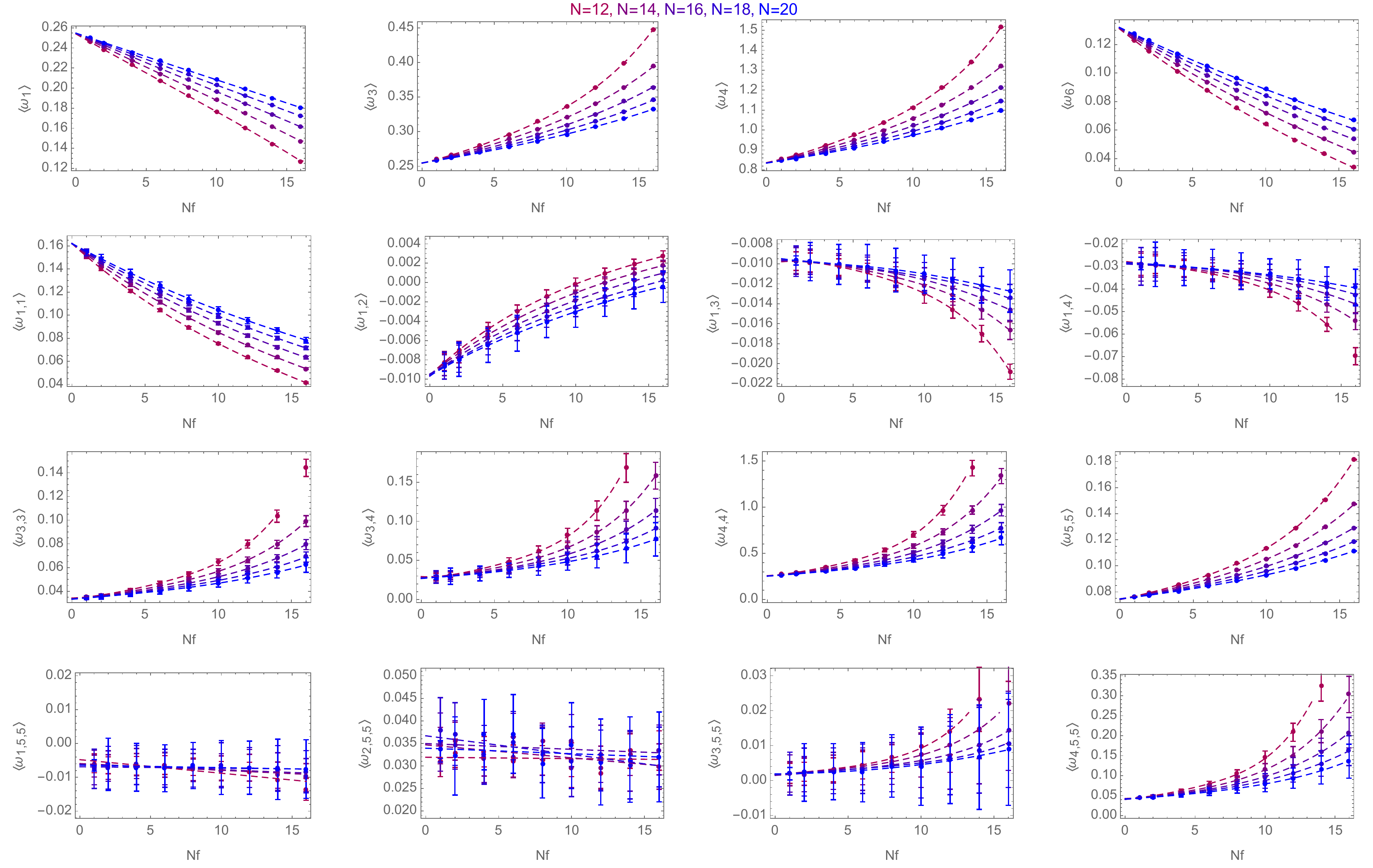}}
\caption{Mean values of the $\omega$'s plotted against $N_f$ for different values of $N$. Dashed lines correspond to either fits of the form $a+bN_f$, $a + b N_f + c N_f^2 + d N_f^3+e N_f^4$ or $a + b N_f + c e^{ d N_f}$.}
\label{fig:b1}
\end{figure}

\section{The high-temperature behaviour of energy $E$,
  Polyakov loop $\langle\kern-1pt P\kern 1pt\rangle$, $\langle R^2\rangle$ and mass susceptibility $\langle{\mathcal C}^m\rangle$ for the supersymmetric model.}
\label{Predictions_for_observables_graphs}
\begin{figure}[H]
\begin{center}
  \makebox[\textwidth]{\includegraphics[width=.9\paperwidth]{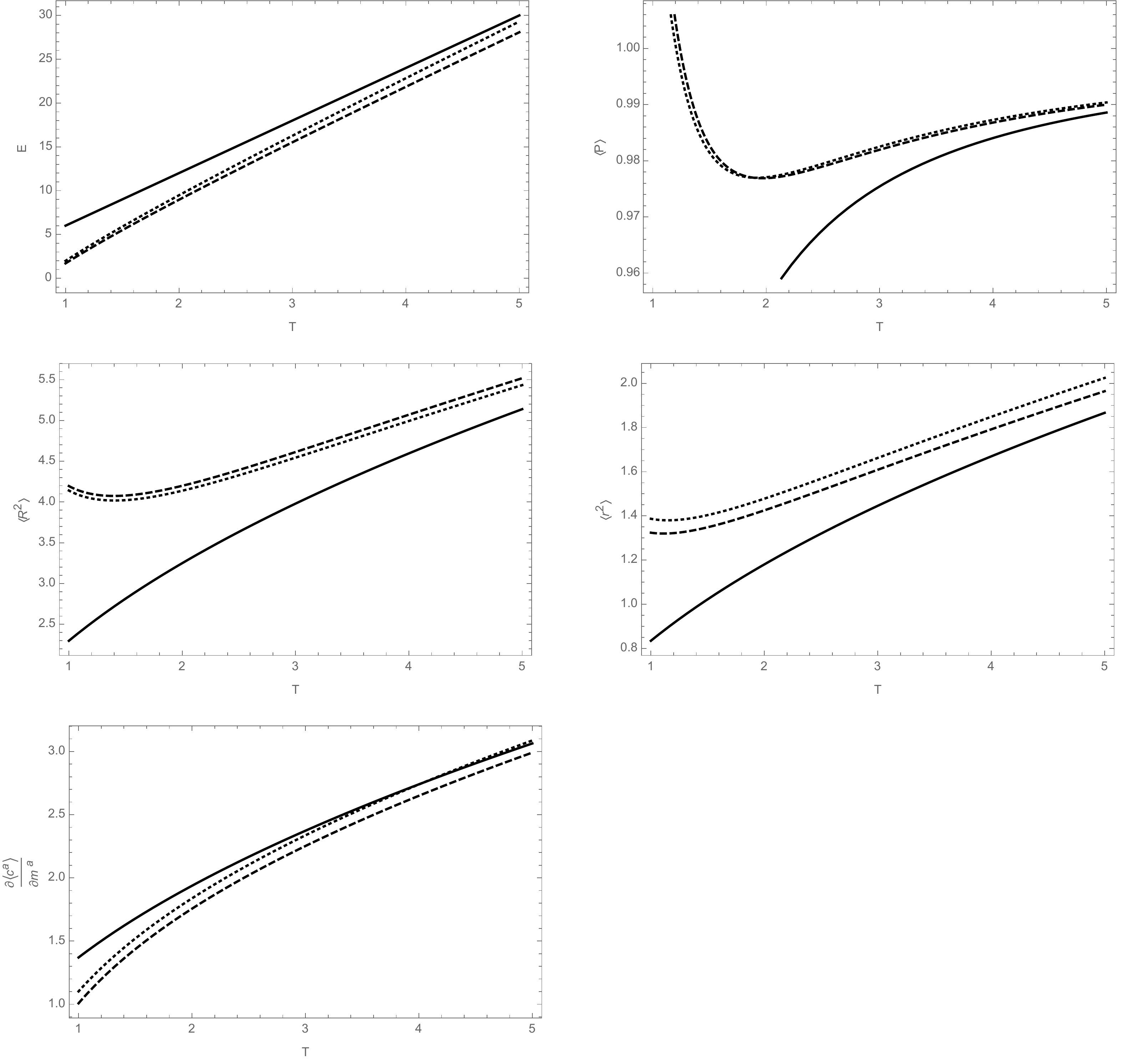}}
\end{center}
\caption{Temperature dependence of physical observables for the
  supersymmetric BD model as defined in \eqref{def of obs} and with the values of $\omega$'s
  from table \ref{fig:tab1}. The solid line is the leading order
  prediction for $N=\infty$, $N_f=1$ while the long dashed line is up to the next to leading order for $N=\infty$, $N_f=1$.
  The third curve with
  short dashes is $N=10$, $N_f=1$. Note that in contrast to the bosonic model
  the high-temperature dependence of the Polyakov loop turns upwards, as
  $T$ decreases,  between $T=1.0$ and $2.0$. This indicates that
  the high-temperature series for $\langle P\rangle$ is not
  reliable in this region.}
\label{fig:c1}
\end{figure}

In this appendix we graphically present the high-temperature predictions for
the BD-model observables the energy $E$, the Polyakov loop $\langle P\rangle$,
the extent of the eigenvalues of the adjoint fields $X^i$ given
by $\langle R^2\rangle$ and the mass susceptibility
$\langle{\mathcal C}^m\rangle$.
Figure \ref{fig:c1} shows the predicted high-temperature
behaviour of the BD-model observables.

\begin{figure}[H]
\begin{center}
  \makebox[\textwidth]{\includegraphics[width=.9\paperwidth]{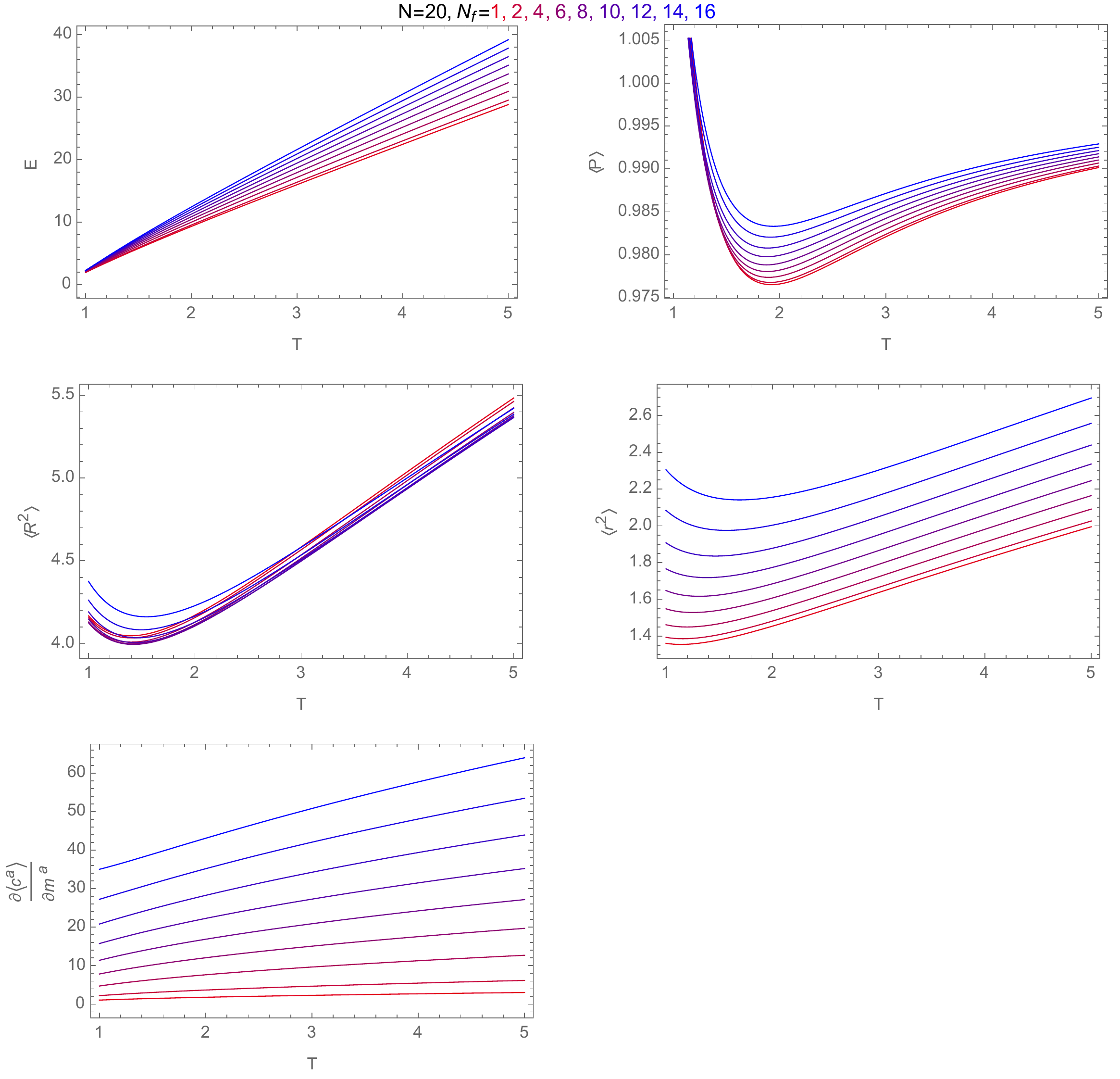}}
\end{center}
\caption{Temperature dependence of physical observables as defined in \eqref{def of obs} with $\omega$'s from table \ref{fig:tab20} for $N=20$ and different values of $N_f$.}
\label{fig:c5}
\end{figure}

\begin{figure}[H]
\begin{center}
  \makebox[\textwidth]{\includegraphics[width=.9\paperwidth]{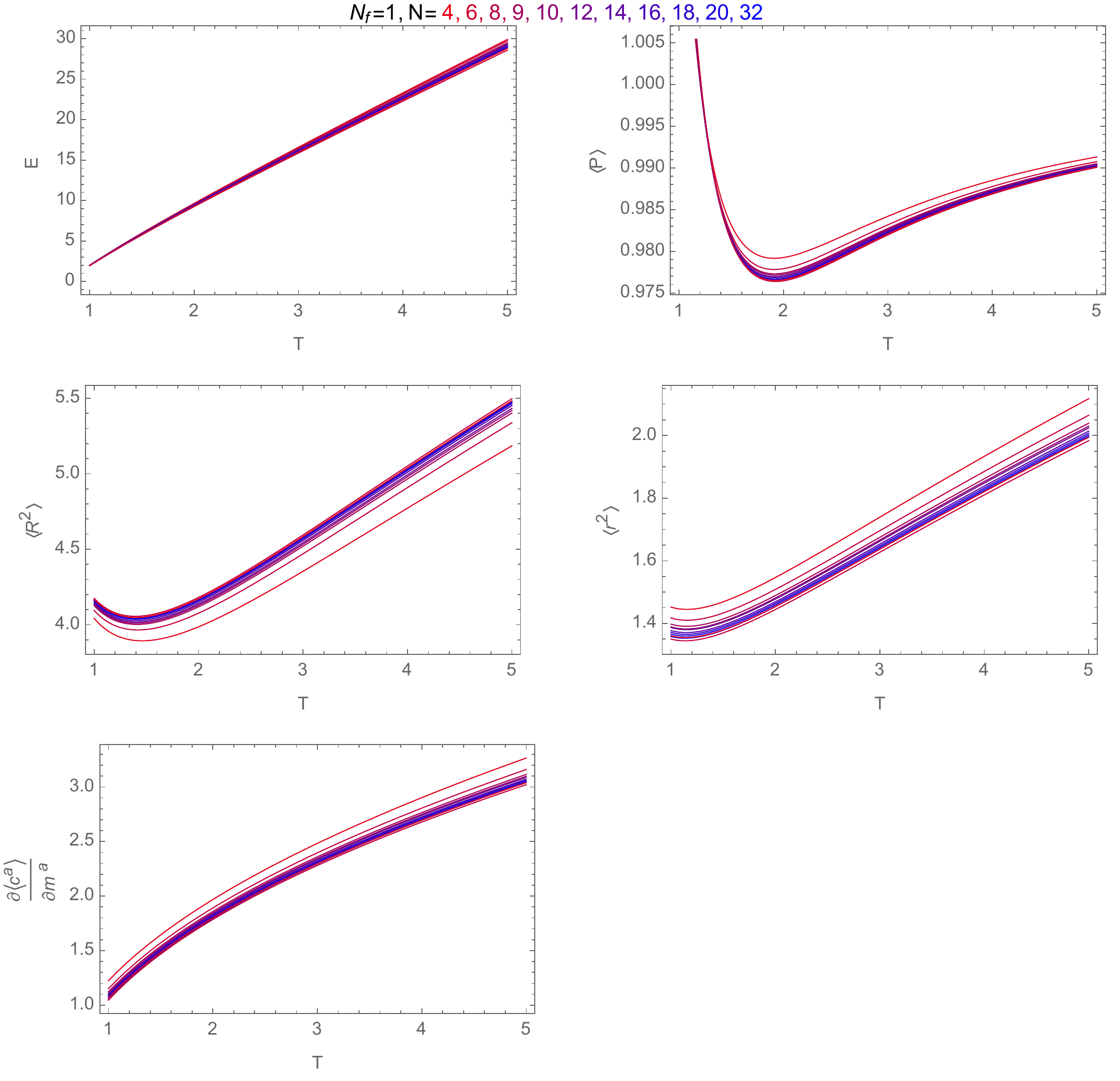}}
\end{center}
\caption{Temperature dependence of physical observables of the
  supersymmetric model as defined in \eqref{def of obs} with $\omega$'s from
  table \ref{fig:tab1} for $N_f=1$ with different values of $N$.}
 \label{fig:c6}
\end{figure}

\begin{figure}[H]
\begin{center}
  \makebox[\textwidth]{\includegraphics[width=.9\paperwidth]{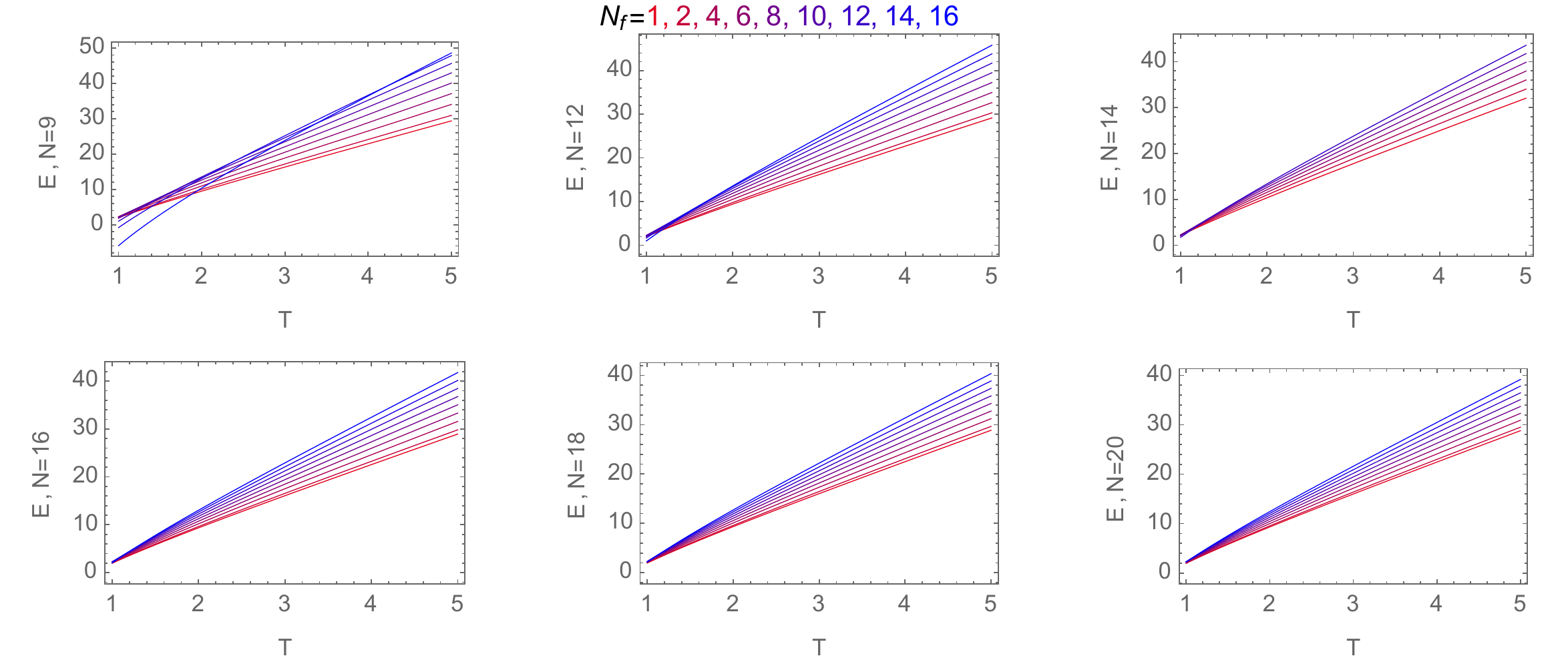}}
\end{center}
\caption{Dependence of the energy on the temperature for the supersymmetric
  model as defined in \eqref{def of obs} for $N=9,12,14,16,18,20$ with different
  values of $N_f$. Note that for each value of $N$ the curves
  (approximately) intersect at a crossing temperature $T_x$. At this point
  the energy is essentially independent of $N_f$.
  Extrapolating the crossing value to large $N$ we find $T_x=0.88\pm 0.02$,
  which is close to the observed transition region of the bosonic BFSS model.}
 \label{fig:d1}
\end{figure}

\vfill\eject

\end{document}